\documentclass[11pt,a4paper]{article}
\pdfoutput=1
\usepackage{jcappub}

\usepackage{amsfonts}
\usepackage{subfigure}
\usepackage{color}
\definecolor{orange}{RGB}{255,127,0}
\usepackage{bm}
\usepackage{url}

\usepackage{chngcntr}
\newcommand{\tref}[1]{\ref{#1}}

\def\gtsima{$\; \buildrel > \over \sim \;$}
\def\ltsima{$\; \buildrel < \over \sim \;$}
\def\gsim{\lower.5ex\hbox{\gtsima}}
\def\lsim{\lower.5ex\hbox{\ltsima}}

\begin{document}
\counterwithout{table}{section}

\title{Secondary antiprotons as a Galactic Dark Matter probe}

\author[a,b]{Carmelo~Evoli}
\author[c]{Daniele Gaggero}
\author[d]{Dario Grasso}

\affiliation[a]{Gran Sasso Science Institute (GSSI)\\ Viale Francesco Crispi 7, 67100 L'Aquila, Italy}
\affiliation[b]{II.~Institut f\"ur Theoretische Physik, Universit\"at
Hamburg\\ Luruper Chaussee 149, D-22761 Hamburg, Germany}
\affiliation[c]{SISSA and INFN, via Bonomea 265, I-34136 Trieste, Italy} 
\affiliation[d]{INFN and Dipartimento di Fisica ``E. Fermi", Pisa University, Largo B. Pontecorvo 3, I-56127 Pisa, Italy} 

\emailAdd{carmelo.evoli@gssi.infn.it}

\date{\today}

\begin{abstract}
{We present a novel determination of the astrophysical uncertainties associated to the secondary antiproton flux originating from cosmic-ray spallation on the interstellar gas. 
We select a set of propagation models compatible with the recent B/C data from PAMELA, and find those providing minimal and maximal antiproton fluxes in different energy ranges. 
We use this result to determine the most conservative bounds on relevant Dark Matter (DM) annihilation channels: We find that the recent claim of a DM interpretation of a gamma-ray excess in the Galactic Center region cannot be ruled out by current antiproton data. 

Finally, we discuss the impact of the recently released preliminary data from AMS-02.
In particular, we provide a reference model compatible with proton, helium and B/C spectra from this experiment. 
Remarkably, the main propagation parameters of this model are in agreement with the best fit presented in our earlier statistical analyses. 
We also show that the antiproton-to-proton ratio does not exhibit any significant anomaly at high energy with respect to our predictions.}
\end{abstract}

\keywords{galactic cosmic rays; antiprotons; dark matter}

\maketitle

\section{Introduction}

The quest for some anomaly in the cosmic-ray (CR) antiproton spectrum has captured the interest of the astro-particle community for more than two decades and is presently one of the main targets of the AMS-02 observatory~\cite{Incagli:2010}. 

The rationale of this effort is the following: If the astrophysical sources, mainly supernova remnants (SNRs), do not inject primary ${\bar p}$ in a relevant amount -- and provided that the secondary antiproton flux due to CR spallation on the diffuse interstellar gas can be accurately computed -- then the measured CR antiproton spectrum becomes a valuable probe of new physics (see, e.g.,~\cite{Lavalle:2012ef} and references therein). In particular, a fascinating scenario is that $\bar{p}$ are copiously yielded by WIMP dark matter (DM) particles annihilating in the dark halo of the Galaxy~\cite{Silk:1984}.

In this context, an important role is played by secondary-over-primary ratios of CR nuclear species: Most noticeably the boron-to-carbon (B/C) ratio is one of the most useful tracers of Galactic CR propagation since boron is expected to be not produced in stars and the cross sections for B production from its main primaries (C and O) are better known.
This channel offers a rather robust constraint on CR models and allows a precise determination of the secondary ${\bar p}$ flux produced in the Inter-Stellar Medium (ISM) 
(see, e.g., \cite{Maurin:2001,Moskalenko02,DiBernardo10,Putze:2010,Trotta:2011})
Moreover, it is crucial to constrain the amount of antiprotons produced and then accelerated in SNRs~\cite{Cholis2013,Mertsch2014}.    

In this paper we make use of the recent measurements of the B/C ratio, together with the proton, helium and light nuclei absolute spectra provided by the PAMELA collaboration~\cite{Adriani:2011cu,Adriani:2014xoa} to constrain the propagation parameters, and to accurately compute the secondary antiproton spectrum in the $0.1 - 10^3$ GeV energy range.  

Our main goal is to achieve a comprehensive evaluation of the relevant uncertainties affecting this computation, extending the work presented in~\cite{Evoli:2011id}.
We focus our attention primarily on: 1) the systematic errors on the antiproton production and inelastic scattering cross sections; 2) the large uncertainty on the propagation parameters in the Galaxy and in the Heliosphere. 

Since several papers have been already published on this subject, we briefly mention here the main novelties of our approach. 
\begin{itemize}

\item We use, for the first time, all the relevant CR measurements taken during the same period and from the same experiment (PAMELA).

\item We use a recent computation~\cite{Kappl:2014hha} of the antiproton production and inelastic cross sections based on the newly available NA49 experimental data~\cite{NA49}.

\item We account for charge dependent solar modulation, including charge sign dependent drifts, by means of the recently developed {\tt Helioprop} code \cite{Maccione:2012cu}.

\end{itemize}

Our approach goes beyond providing an updated range of allowed propagation setups to be used to propagate dark matter annihilation products. 
The widely used MIN-MED-MAX propagation models are defined as those models giving the maximal, median, and minimal supersymmetric (primary) antiproton flux and are compatible with B/C analysis \cite{Donato:2003}. 
Such set of models, however, cannot be confused, as often done in the literature (see however \cite{Donato:2008jk,Bringmann:2006im}), with those providing the allowed range of secondary antiprotons. 
Indeed, while the primary antiproton flux is strongly dependent on the diffusive halo scale height and on the physical conditions in the Galactic center region, once the propagation parameters are set to fit the B/C, the secondary flux is almost independent on those quantities. 
Since the models providing the maximal/minimal secondary antiproton flux are dependent on the antiproton energy, we obtain for each energy bin the allowed flux compatible with CR nuclear measurements and the  cross section uncertainties. We then constrain DM models using the minimum secondary antiproton flux evaluated in that way as background.

Our results allow us to determine up-to-date conservative constraints on a representative sample of DM models and to compare these limits with the recent claim of a gamma-ray excess in the GC associated with DM annihilations in that region.

In the last section, the preliminary results recently released by AMS-02 up to $450$~GeV/n \cite{AMS_preliminary} are considered as well. 
In particular, we investigate the presence of an excess in the $\bar{p}/p$ spectrum at energies higher than previously measured by PAMELA.
Given that the AMS-02 B/C data are still preliminary, we do not aim at providing a dedicated statistical analysis. 
On the other hand, we select a single model providing the best fit of AMS-02 B/C data, and we tune the injection spectra to reproduce the proton and Helium data from the same experiment. Finally, we compare the predicted antiproton fluxes for this model with the new measurements. We discuss these results in the last Section.
  
\section{CRs in the Galaxy}

\subsection{CR propagation in the ISM}\label{sec:prop}

We model CR propagation in the Galaxy using the publicly available numerical code {\tt DRAGON}~\cite{Evoli:2008dv,Dragonweb}. 
This code solves the diffusion-convection transport equation for all CR species in the Galaxy taking into account all the relevant processes, including energy losses and spallation due to the interaction with the ISM. 
We adopt the nuclear cross-section database taken from the public version of {\tt GALPROP}~\footnote{\url{http://galprop.stanford.edu}}~\cite{Moskalenko1998}.

We assume cylindrical symmetry in the Galaxy. The propagation region is a classical cylindrical box with radial extension $R = 20$~kpc and height ($L$) varying in the range ($2-16$) kpc. 
The lower limit of the halo size is determined by the observed diffuse synchrotron emission and by comparison between the computed low-energy secondary positron spectrum and PAMELA data \cite{DiBernardo:2012zu,Lavalle:2014kca}. 
The upper limit is chosen in agreement with the available measurements of the $^{10}$Be/$^9$Be ratio. 
In the next section we will show that secondary antiprotons from CRs are almost unchanged by varying $L$ in the given range, making the choice of the allowed domain not crucial in this context. 
We optimize the spatial resolution, as well as the time steps involved in the numerical algorithm, in order to evaluate the $\chi^2$ against PAMELA data with a precision $< 1$\%.

We adopt a two-dimensional ($r$ and $z$) propagation configuration since it is commonly used to study secondary CR antiproton production and also to keep the computational time short enough to scan over a large number of models.
A more realistic three-dimensional approach could give a more accurate prediction of the CR fluxes, by taking the spiral arm distribution of CR sources and of the ISM gas into account~\cite{Gaggero2013,Kissmann:2015}.
However we confirmed that these differences are much smaller than the other uncertainties under investigation in this work. 

We adopt the following standard expression for the diffusion coefficient $D$:

\begin{equation}
\label{eq:diff_coeff}
D(\rho) = D_0 ~\beta^\eta  \left(\frac{\rho}{\rho_0}\right)^\delta \;,
\end{equation}
where $\rho \equiv p/Z$ is the rigidity of the nucleus characterized by charge $Z$ and momentum $p$, and $\eta$ parametrerizes the low-rigidity behavior of $D$.
While kinetic quasilinear theory predicts $\eta=1$ for the dependence of the diffusion coefficient on the particle speed, several effects may give rise to a different effective behavior (see e.g. \cite{Ptuskin06,EvoliYan2014}). 
Bearing these considerations in mind, we leave $\eta$ as a free parameter to be fitted against low-energy CR data. 

Diffusive reacceleration is parameterized in terms of the Alfv\'en velocity $v_A$ and we assume that this process takes place in the entire propagation region.
In fact, this circumstance is expected if CRs are responsible for generating their own turbulence, e.g.,~by CR-induced streaming instability~\cite{Cesarsky,Blasi:2012yr}.  
Our approach is then different from semi-analytical methods where reacceleration is active only in the Galactic disk with height $\sim 100$~pc (see, e.g.,~\cite{Donato01}). 

We also account for a convective wind with velocity $V_C(z)$ vanishing at $z = 0$ and growing linearly with $z$ with the gradient $dV_c/dz$. 
Although $dV_c/dz$ can be as large as $\sim 100$~km s$^{-1}$ kpc$^{-1}$ in the inner few kpcs of the Galactic disk \cite{Everett,Breitschwerdt}, the observables we are considering in the present analysis do not probe that region. On the other hand, the local convective velocity is severely constrained by $^{10}$Be/$^9$Be measurements \cite{Strong:2007nh} and, for this reason, we require $dV_c/dz$ to be less then $10$~km s$^{-1}$ kpc$^{-1}$.
By making this choice we introduce another difference with respect to semi-analytical approaches where a constant convection velocity is adopted~(see e.g.~\cite{Donato01}). 

For the CR astrophysical source distribution we consider the following form:
\begin{equation}
Q_{i}(E_{k},r,z) =  f_{\rm SNR}(r,z)\  q_{0,i}\ \left(\frac{\rho(E_{k})}{\rho_0}\right)^{- \gamma_{i}} \;,
\label{eq:source}
\end{equation}

Our spatial source distribution ($f_{\rm SNR}$) follows the Galactic supernova remnants profile inferred from pulsars and stellar catalogues as given in \cite{Ferriere:2001rg}. 
We allow different source spectral indexes for the different nuclear species ($\gamma_{i}$), and we fix their values -- together with the species relative abundances ($q_{0,i}$) -- against recent experimental data. 

Proton and helium spectra exhibit a change of slope at a rigidity between $\sim 200$ and $300~{\rm GV}$ \cite{Adriani:2011cu}: We model it by assuming a spectral break in the injection slopes of these nuclei. These features have relevant implications for the secondary antiproton flux.
An alternative possibility for the origin of this break is a change in the ISM turbulence power spectrum and hence in the diffusion coefficient. In this context, high-energy CR antiprotons could be successfully used to discriminate among the two interpretations~\cite{Evoli:2011id}, but in the energy range where antiprotons are currently measured the two scenarios are quantitatively equivalent.

\subsection{Secondary antiproton production in the ISM}

The source term of secondary antiprotons produced from the interaction of CR nuclei with the ISM is given by the convolution of the antiproton production differential cross-section, $d\sigma/dE_k$, and the interstellar CR nuclei differential energy flux, $d\Phi/dE_k$.

The differential $\bar{p}$ production rate per volume and energy takes the form:
\begin{equation}
Q_{\bar{p}} (E_k) = \sum\limits_{i={\rm H},{\rm He}} \sum\limits_{j={\rm H},{\rm He}} 4 \pi \int_{E_k^{\rm th}}^\infty d E'_k \left( \frac{d\sigma}{dE_k} \right)_{ij} n_i \phi_j (E'_k)  
\end{equation}

where $E'_k$ and $E_k$ are the kinetic energies per nucleon of the incoming nucleus (with threshold energy $E_{\rm th} = 6 m_p$) and the outgoing antiproton, respectively.
$n_i$ denotes the interstellar gas density.

Existing parameterizations of the antiproton production cross-sections \cite{1982PhRvD..26.1179T} are mainly based on experimental data earlier than 1980. Recently, two works (\cite{Kappl:2014hha,diMauro2014}) have improved these old computations making use of new precision data from the NA49 experiment at CERN~\cite{NA49}.
In particular, in \cite{Kappl:2014hha} the authors extended previous calculations by including the new laboratory measurements and a careful treatment of antiprotons arising from antineutron and hyperon decay.
By using a revised approach to proton-helium scattering, the authors of \cite{Kappl:2014hha} are also able to compute the production cross-sections for the processes: $p + p \rightarrow \bar{p} X$, $p + He \rightarrow \bar{p} X$, $He + p \rightarrow \bar{p} X$ e $He + He \rightarrow \bar{p} X$, where the first particle is the impinging primary CR while the second one is the target interstellar material. 

We make use of their results for the antiproton cross sections, as well as their estimate of the related errors, in order to evaluate the nuclear uncertainties on our prediction of the secondary antiproton fluxes in the Galaxy.
Finally, we also checked that our conclusion are unchanged by adopting the parameterization for the $p+p$ scattering proposed by~\cite{diMauro2014} and based on the full set of available measurements.
  
\subsection{Antiprotons from dark matter annihilations}\label{sec:dm}

Other than spallation of CRs on the ISM nuclei, antiprotons can be produced in the Galaxy by DM in pair annihilation or decay events.
Antiproton measurements provide an invaluable tool to constrain such primary contribution, since the ratio between DM signal and background from standard astrophysical sources is usually much larger in the antiproton channel with respect to all other indirect detection methods. 

We compute the DM contribution to the antiproton flux as described in several previous works, e.g.,~\cite{Evoli:2011id,Cirelli2014}.
In brief, we assume that the source function ($Q_{\rm DM}$) for the WIMP DM component scales with the DM particle number density times the probability of annihilation and the antiproton yield per annihilation:
\begin{equation}
Q_{\rm DM} (E_k,r,z) \,=\, \frac{1}{2} \, \frac{\rho^2_{\rm DM}(x)}{m^2_{\rm DM}} \, \langle \sigma v \rangle \, \frac{d N_{\bar{p}}}{d E_k} (E_k)
\end{equation}
where $\langle \sigma v \rangle$ is the thermally averaged annihilation cross section and $\rho_{\rm DM}(x)$ is the DM density profile as function of the galactocentric distance $x = \sqrt{r^2+z^2}$.

The DM profile is only poorly constrained by direct observations and its shape is usually inferred from N-body simulations of gravitational clustering. 
In our analysis, we adopt two spherically symmetric profiles: a standard Navarro-Frenk-White (NFW) \cite{NFW1996}, and a generalized NFW (gNFW) as defined, e.g., in~\cite{Cirelli2014}. 
The free parameters for the NFW profile are chosen following the analysis in \cite{Catena2012}, while for the gNFW profile we adopt a contracted profile ($\gamma = 1.2$) as suggested by the recent claim of a DM associated excess in the GC region~\cite{Calore2014}.

For the choice of the annihilation channels, we focus here on two sample cases which have been recently investigated in connection to hints of DM signals (both direct and indirect), but potentially giving a sizable antiproton flux as well~\cite{Evoli:2011id}.
In particular, we consider as primary annihilation channels: ${\rm DM \, \, DM} \rightarrow b \bar{b}$ and ${\rm DM \, \, DM} \rightarrow W^+W^-$, and we take the corresponding antiproton yields from the {\tt PPPC4DMID} \cite{Cirelli2010,Ciafaloni2010}.  
  
\subsection{Propagation in the Heliosphere}\label{solarmod}

Low-energy ($\lesssim$ 10 GeV) charged CRs are influenced by the solar magnetic field in the final stage of their propagation process.
The main effect of the interaction with the heliosphere is a momentum decrease and the consequent alteration of the interstellar spectrum.

This process is usually described in the context of the ``force field approximation''~\cite{Gleeson68}. 
According to this model, the energy spectrum $d\Phi_{\rm TOA}/dE_{k, \rm TOA}$ of particles reaching the top of the atmosphere (TOA) with kinetic energy $E_{k, \rm TOA}$ and momentum $p_{\rm TOA}$ is related to the local interstellar spectrum (LIS) $d\Phi_{\rm LIS}/{dE_{k, \rm LIS}}$ as it follows
\begin{equation}\label{eq:Fisk}
\frac{d\Phi_{\rm TOA}}{dE_{k, \rm TOA}} \,=\, \frac{p_{\rm TOA}^2}{p_{\rm LIS}^2} \, \frac{d\Phi_{\rm LIS}}{dE_{k, \rm LIS}},\qquad
E_{k, \rm LIS} = E_{k, \rm TOA} + |Ze| \phi, 
\end{equation}
where $\phi$ is the Fisk potential and parameterizes the kinetic energy losses.
Within this approach the effects of modulation for protons and antiprotons are identical.

More realistic drift models, however, predict a clear charge sign dependence for the heliospheric modulation of positively and negatively charged particles.
Moreover, the modulation effect depends on the polarity of the solar magnetic field (SMF), which changes periodically every $\sim 11$~years.
The SMF is observed with opposite polarities in the northern and southern hemispheres. At the interface between opposite polarity regions a heliospheric current sheet (HCS) is formed.
The phenomenologically parameter $\alpha$, known as the ``tilt angle'', sizes the angular extension of the HCS oscillations. 
The magnitude of $\alpha$ depends on the solar activity and has a large influence on the intensity of the modulation.

A more accurate description then may be achieved by means of dedicated numerical simulations of the CR transport in realistic models for the interplanetary magnetic field.

In order to assess the impact of realistic model for solar modulation, we make use of the recently developed {\tt HelioProp} code~\cite{Maccione:2012cu}. 
This code solves the equation describing CR transport in the heliosphere by means of the stochastic differential equation method \cite{Strauss:2011,Strauss:2012}.  
To characterize the model for solar propagation we have to specify the solar magnetic field geometry, the properties of diffusion, and those of winds and drifts.

For the interplanetary diffusion tensor we assume $H \propto {\rm diag}({\lambda}_{\|}, {\lambda}_{\perp, r}, {\lambda}_{\perp, \theta})$, where the parallel and perpendicular components are set with respect to the direction of the local magnetic field. 
The CR mean free path parallel to the magnetic field as function of the particle rigidity $\rho$ is parameterized as $\lambda_{\|} = \lambda_0 (\rho / 1 {\rm GeV})^\delta (B/B_{\rm sun})^{-1}$, where $B$ is the magnetic field and $B_{\rm sun}$ is its normalization value at the Earth position, for which we adopt $B_{\rm sun} = 5$ nT. 
For the perpendicular diffusion we assume ${\lambda}_{\perp, r} = {\lambda}_{\perp, \theta} = 0.02 \lambda_{\|}$, as results from numerical simulations~\cite{Giacalone1999}.

Finally, for modeling the magnetic field geometry, winds and drifts associated to the antisymmetric components of the diffusion tensor we follow~\cite{Maccione:2012cu}. 

\section{Method and results}

\begin{figure}[t]
\begin{center}
\includegraphics[width=0.49\textwidth]{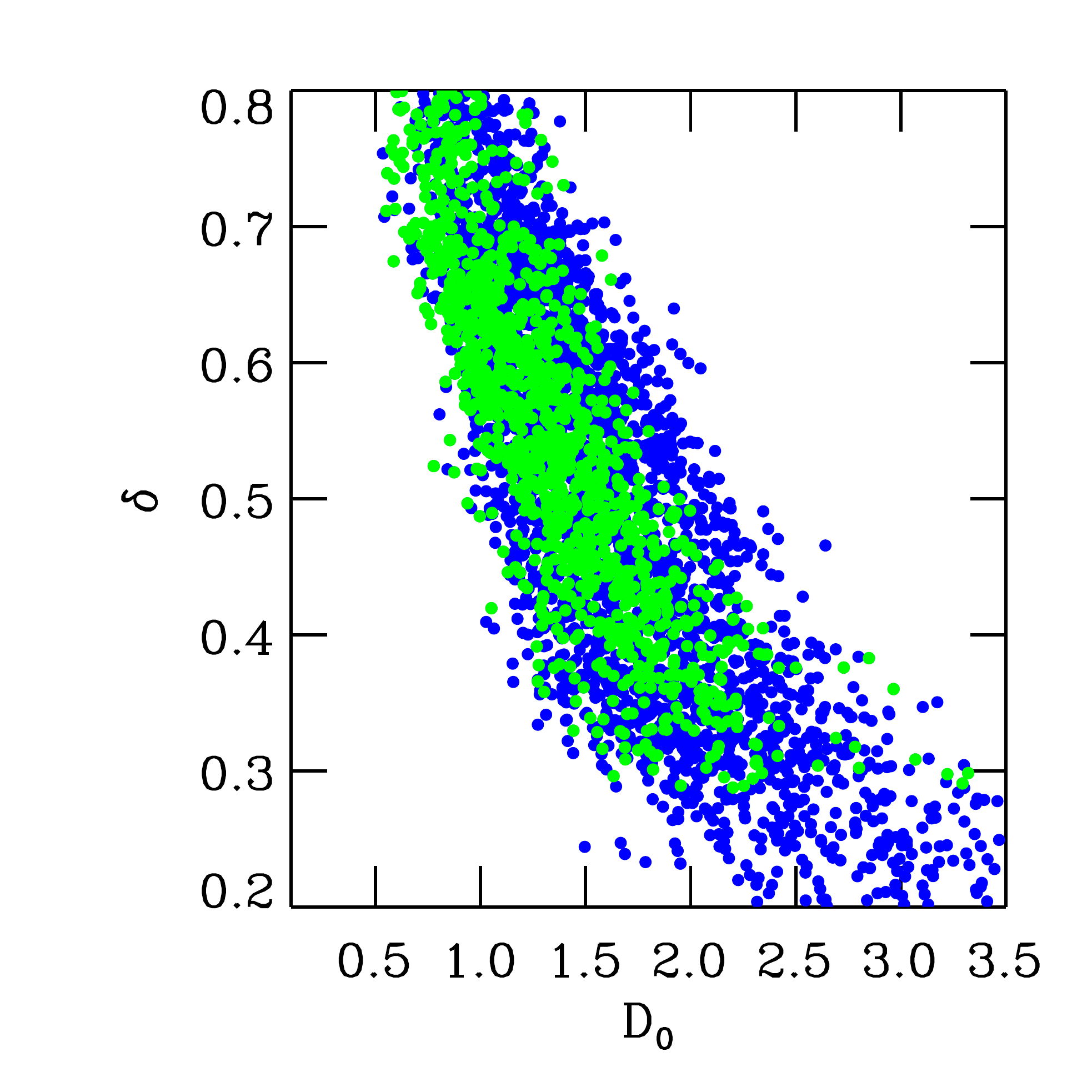} \hspace{\stretch{1}}
\includegraphics[width=0.49\textwidth]{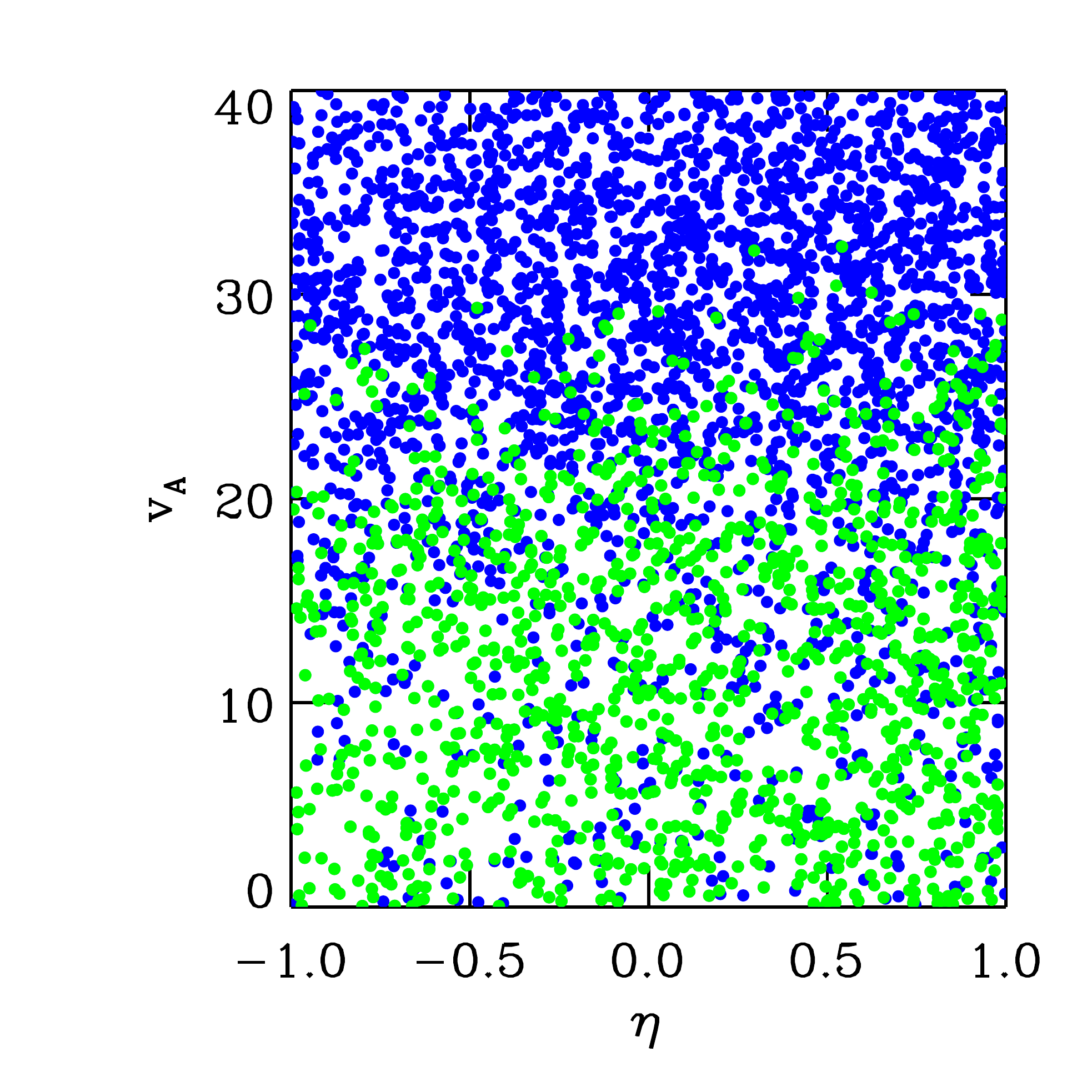}
\caption{The good models as discussed in section~\ref{sec:selection} in the $\delta - D_0$ (left) and $v_A - \eta$ (right) planes. Blue points correspond to models compatible with the PAMELA B/C ratio. Green points show the good models obtained using the PAMELA primary proton in addition.}
\label{fig:dvsdelta}
\end{center}
\end{figure}

\begin{figure}[!t]
\begin{center}
\includegraphics[width=0.5\textwidth]{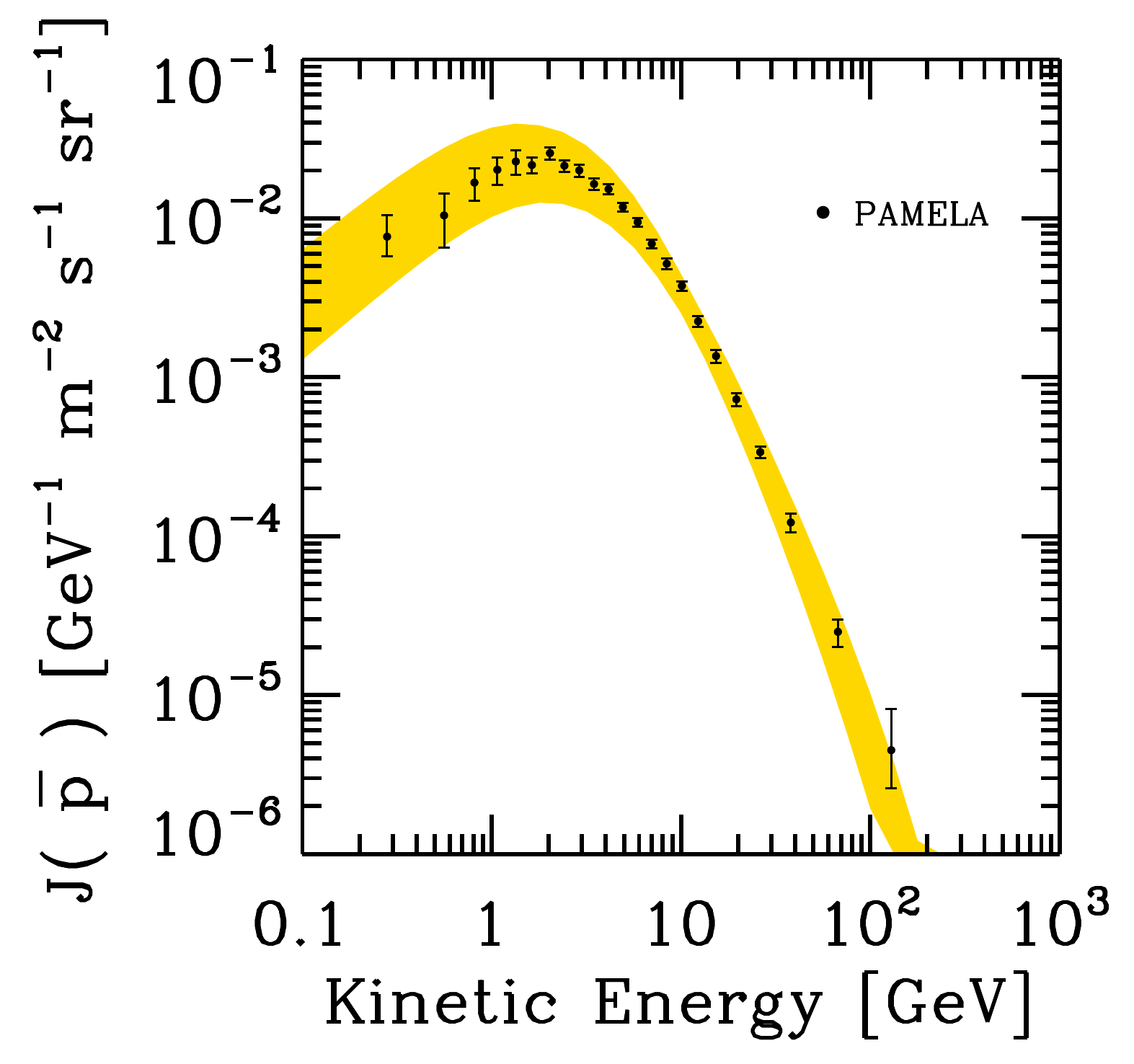}
\caption{The envelope of the secondary antiproton spectra computed with the different propagation models found to reproduce the B/C and primary spectra.}
\label{fig:minmax}
\end{center}
\end{figure}

\begin{figure}[!t]
\begin{center}
\includegraphics[width=0.45\textwidth]{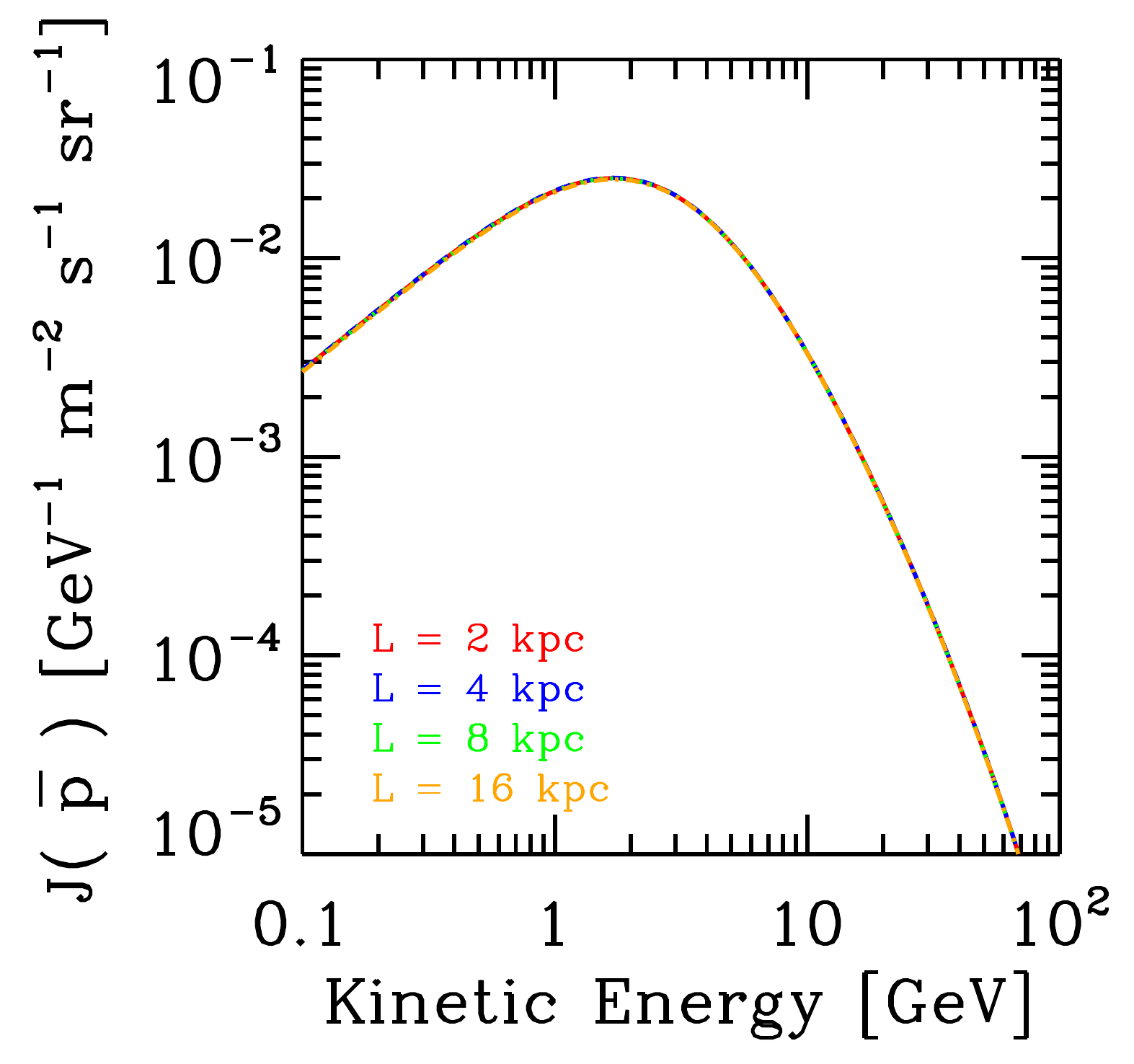}
\includegraphics[width=0.45\textwidth]{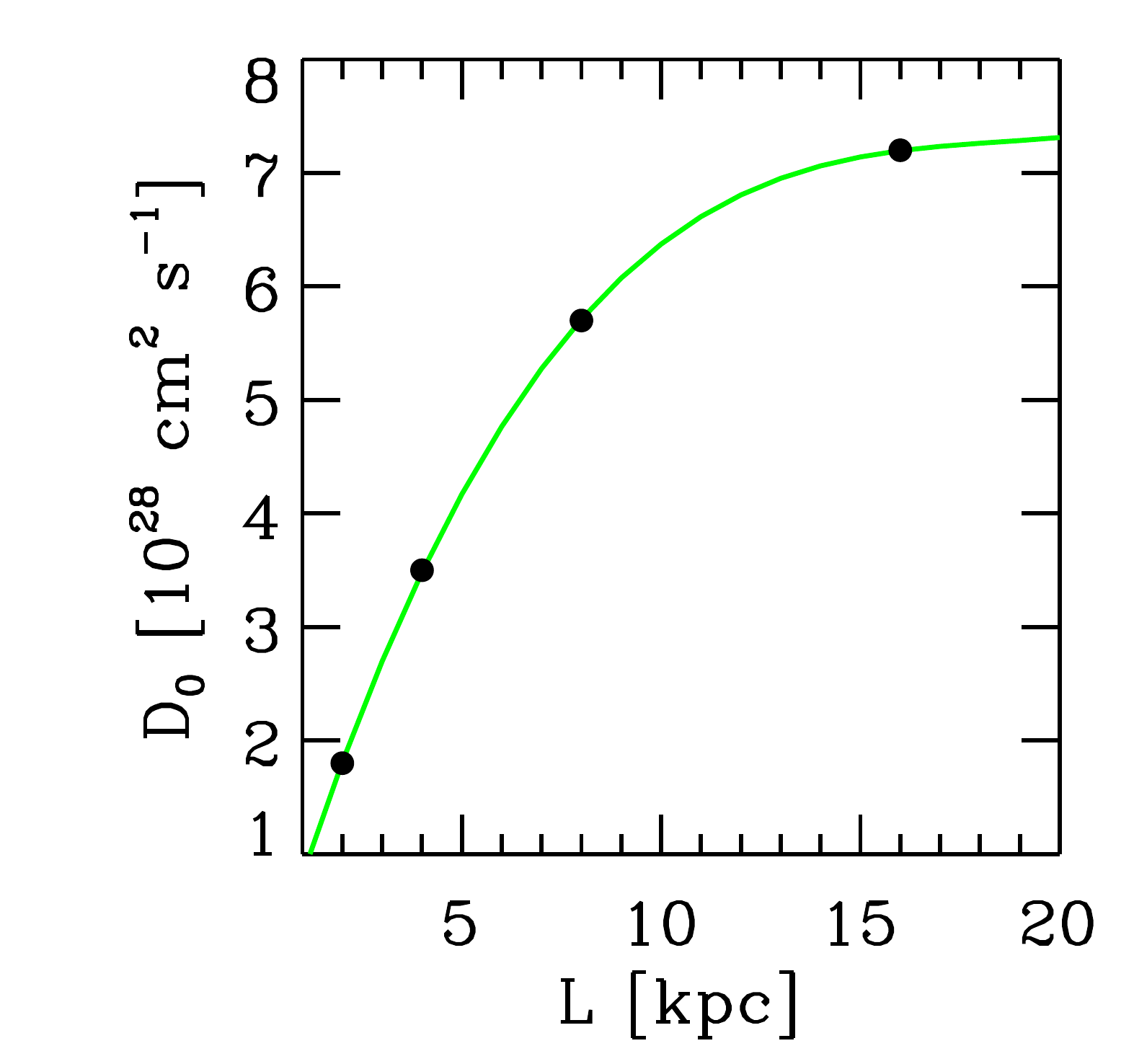}
\caption{{\it Left plot:} The secondary antiproton flux computed for different values of the halo height $L$. 
{\it Right plot:} The normalization of the diffusion coefficient required, for each value of $L$, to reproduce the B/C ratio (the green line is used to guide the eye).}
\label{fig:vsL}
\end{center}
\end{figure}

\subsection{Selection of CR propagation models}\label{sec:selection}

We focus our attention on the free parameters of our propagation model (see section~\ref{sec:prop}) within the range given in table~\tref{tab:parameters}.
We notice that the observables we are considering in our analysis are not sensitive to $L$, we then use a fixed value $L=2$~kpc and we evaluate the impact of different values for this parameter afterwards. 

\begin{table}[h]
\begin{center}
\begin{tabular} {|c|c|c|c|}
\hline
\hline
Parameter & Min value & Max value & Units \\
\hline
$\eta$     & $-1$ & $1$ & \\
$D_0$ & $0.1$ & $10$ & $10^{28}$~cm$^2$/s \\ 
$\delta$ & $0.2$ & $0.8$ & \\ 
$dV_c/dz$ & $0$ & $10$ & km/s/kpc \\ 
$V_A$ & $0$ & $100$ & km/s \\
\hline
\hline
\end{tabular} 
\caption{The free parameters of our propagation model with their associated range. {The parameter $L$ is taken fixed to $2$ kpc.}}
\label{tab:parameters}   
\end{center}
\end{table}

For each model selected randomly in the parameter space described above, we fix the injection slopes and the source abundances for primary nuclei heavier than carbon by fitting CREAM data above 10 GeV/n~\cite{Cream2014}. 
For the carbon, helium and proton parameters, including the Fisk potential $\phi$, we fit the recent PAMELA Carbon~\cite{Adriani:2014xoa} and proton~\cite{Adriani:2011cu} measurements at energy below break at $\sim 200$~GV.
We assume Boron is entirely secondary. It was shown by~\cite{Tomassetti:2012} that, neglecting the production and acceleration of secondary nuclei inside SNRs, the $\delta$ may be underestimated by a factor of $\sim 5 - 15$\% (see also~\cite{Genolini:2015}).
We checked that the Fisk potential gives an accurate description of modulated spectra compared against the more realistic predictions provided by the {\tt Helioprop} simulations. 
Using a charge-dependent formalism for the modulation is relevant only when we compare differently charged particle spectra. We discuss this in detail in section~\ref{sec:charged}.

With the given set of diffusion and source parameters we are now able to calculate the B/C ratio.

We identify a model as a \emph{good} one, if it reproduces the $B/C$ data as well as proton and carbon data within the 3$\sigma$ limits.

In particular, we compute the $\chi^2$'s against $B/C$ ($\chi^2_{BC}$), proton ($\chi^2_p$) and carbon ($\chi^2_C$) measurements for each propagation model and we accept the model if each $\chi^2$ is smaller than the corresponding threshold ($\chi^2_{3\sigma}$) reported in table~\tref{tab:chi2}. 
The $\chi^2_{3\sigma}$ values reported in the table are calculated according to the $\chi^2$ distribution function with a number of degrees of freedom $\mathcal{F}= \mathcal{D} - \mathcal{P}$ with $\mathcal{D}$ equals to the number of data points and $\mathcal{P}$ to the number of parameters. 
The $3\sigma$ threshold implies a $99.73$\% probability to get a smaller $\chi^2$ value or, equivalently, a p-value of $0.27$\%.

A similar strategy (using only $\chi^2_{BC}$) has been adopted in~\cite{Donato01} to select the models in order to determine the well-known MIN, MED and MAX setups.

\begin{table}[t]
\begin{center}
\begin{tabular} {|c|c|c|c|}
\hline
\hline
Observable & Data points $\mathcal{D}$ & \# of parameters $\mathcal{P}$ & $\chi^2_{3 \sigma}$ \\
\hline 
B/C & 18 & 5 & 29.79 \\ 
Proton flux & 71 & 7 & 98.21 \\
Carbon flux & 16 & 7 & 21.58 \\
\hline
\hline
\end{tabular} 
\caption{$\chi^2$ values corresponding to 3$\sigma$ for the different observables.}
\label{tab:chi2}   
\end{center}
\end{table}

We repeat the procedure until $N=10^4$ good models are selected.

In figure~\ref{fig:dvsdelta} we show where the selected models are located in the $\delta$-$D_0$ and $v_A$-$\eta$ planes.
We find that PAMELA data allow $\delta$ to vary between $0.2$ and $0.8$ and this parameter is strongly anti-correlated with $D_0$. { This anti-correlation can be explained as follows: An increase of $\delta$ corresponds to a lower secondary-to-primary ratio at high energies, therefore $D_0$ must be smaller in order to enhance the high-energy secondary production.}
The low-energy parameter $\eta$ is strongly degenerate with the solar modulation potential and, as a consequence, the available data do not constrain $\eta$ within the chosen range.
By contrast, the Alfv\'en speed parameter ($v_A$) is found to be bounded by $\sim 30$ km$/$s if  primary spectra are included in the analysis and no low-energy breaks in the injection slopes are admitted. 

The propagation model giving the best fit of the PAMELA B/C and proton data is characterized by the following parameters: $D_0 = 1.6$, $\delta = 0.41$, $v_A = 8.5$, $dV_C/dz = 1.6$, $\gamma_C = 2.56$, $\gamma_H = 2.47$.

\subsection{Extreme models determination}

In order to evaluate the propagation uncertainty in the determination of the secondary antiproton flux, we show in figure~\ref{fig:minmax} the envelope of the secondary antiproton spectra computed with the propagation models selected beforehand.
At lower energies the uncertainty band widens since more parameters are necessary to model CR propagation, while at larger energies the main uncertainty comes from the poor determination of the $\delta$ parameter. 
Our result is in agreement with~\cite{Bringmann:2006im}.

We notice that the propagation model giving the best minimum (maximum) flux of secondary antiprotons is not univocally determined over the entire energy range $0.1 - 100$~GeV. 
In fact, at the different energies the minimum (maximum) flux is associated with a different propagation setup.

To give an example of the different models selected by varying the energy at which we evaluate the extreme fluxes, we provide in table~$3$ the model parameters associated with the minimum and maximum antiproton flux at $1$, $10$ and $100$ GeV.

\begin{table}
\begin{center}
\begin{tabular}{ |cccccccccc| }
\hline
\hline
E$_k$ & $\eta$ & D$_0$ & $\delta$ & v$_A$ & dV$_C/$dz & $\gamma_p$ & q$_C$                & $\gamma_C$ & $\phi$ \\ 
$[$GeV$]$  &            & units    &               & $[$km/s$]$ & $[$km/s/kpc$]$   &                       & $[\times 10^3]$  &                        & $[$GV$]$ \\
\hline
\multicolumn{10}{ |c| }{Min models} \\
\hline 
1      & 0.30  & 3.32   & 0.30 & 32.2   & 0.04 &  2.58 & 2.74 & 2.53 & 0.77 \\
10    &  0.68 &  2.85  & 0.38 & 28.6   & 0.03 &  2.54 & 2.83 & 2.48 & 0.86 \\
100  & -0.16 & 1.17 & 0.75 & 9.31 & 6.78 & 2.38 & 3.40 & 2.17 & 0.57 \\
\hline
\multicolumn{10}{ |c| }{Max models} \\
\hline
1     & 0.84 & 0.85 & 0.74 & 0.52 & 5.65 & 2.40 & 3.80 & 2.18 & 0.53 \\
10   & -0.92 &  0.83 &  0.68 &  7.71 &  4.05 &  2.44 & 4.03  & 2.22 & 0.54 \\
100 & 0.60  &  2.85 &  0.23 &  27.4 &  6.88 &  2.62 & 2.95  & 2.59 & 0.75  \\
\hline
\hline
\end{tabular}
\label{tab:minmax}
\caption{Model parameters giving the minimum (maximum) contribution of secondary anti-protons at energy $E=1$, $10$, $100$~GeV.}
\end{center}
\end{table}

Some trends emerge from this comparison.
The minimal models selected at higher energies feature a larger $\delta$, while high reacceleration reduces the antiproton flux at low energies. 
As expected the maximal models show the opposite behavior, indeed the hardest value allowed for $\delta$ gives the maximum contribution to the antiproton flux at higher energies. 

It is important to remark here that our poor knowledge of the halo size does not affect these conclusions. 
To show this, we select the propagation model giving the B/C best fit and we test different values for $L$ up to $16$~kpc. 
In order not to lose the perfect agreement with the secondary over primary data, we increase the $D_0$ value accordingly (see the right plot in figure~\ref{fig:vsL}). 
As shown in figure~\ref{fig:vsL}, different choices for $L$ in this range do not affect our predictions for the secondary antiproton flux.

Although in this paper we assumed a uniform value of $\delta$ in the whole Galaxy, it was recently shown that diffuse $\gamma$-ray data favor a scenario characterized by radially-dependent CR transport properties~\cite{Gaggero:2014,Gaggero:2015}.
In order to investigate the possible impact of that scenario on our results, we computed the local secondary antiproton spectrum for the KRA$_\gamma$ model considered in those papers finding a negligible correction. 

\subsection{Antiproton production cross-section uncertainties}

We compare here the propagation uncertainties derived in the previous sections with those associated with the antiproton production processes.

\begin{figure}[!t]
\begin{center}
\includegraphics[width=0.5\textwidth]{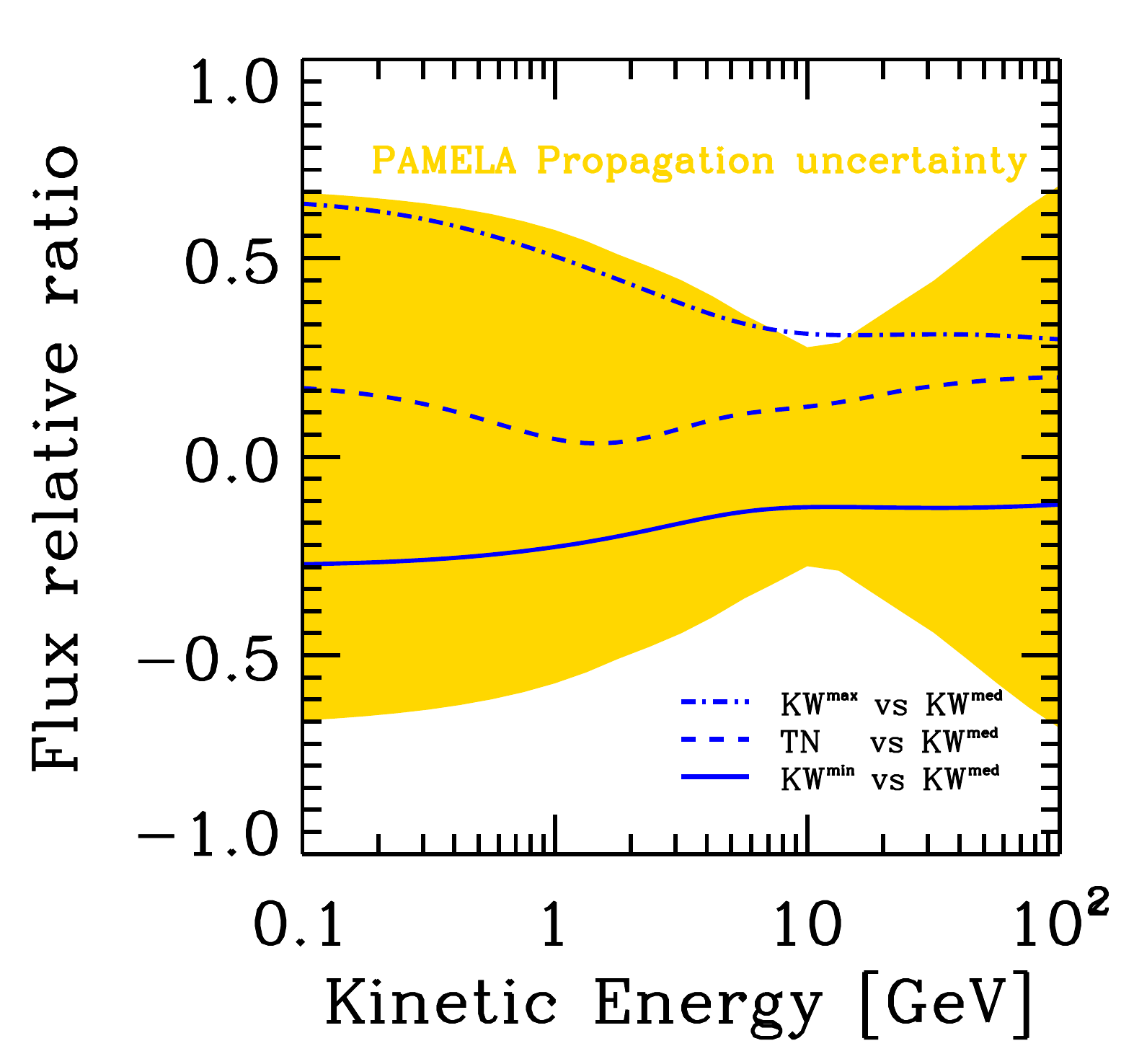}
\caption{Comparison between propagation and nuclear uncertainties. {\it Yellow band:} Error on the $\bar{p}$ flux due to the uncertainty in the propagation parameters. {\it Blue lines:} The {relative difference between the $\bar{p}$ flux computed using the fiducial cross section from~\cite{Kappl:2014hha} and:  (dot-dashed) the maximal model from~\cite{Kappl:2014hha}, (solid) the minimal model from~\cite{Kappl:2014hha}, (dashed) the parameterization from~\cite{1982PhRvD..26.1179T,Moskalenko1998}.} }
\label{fig:cs}
\end{center}
\end{figure}

In figure~\ref{fig:cs}, we show the {relative difference} between the minimum (maximum) secondary antiproton flux and that obtained using the best-fit propagation model. 
The corresponding region represents the uncertainty on the secondary flux associated with galactic propagation.

We compare this uncertainty band with the relative differences associated with production cross sections.
To this end, we compute secondary antiprotons with the new prescriptions recently proposed by~\cite{Kappl:2014hha} and we evaluate them against the traditional fitting relations given in~\cite{1982PhRvD..26.1179T,Moskalenko1998}.

We find that nuclear uncertainties can be as large as $50$\% even at $\sim 100$ GeV, and are much larger below few GeVs.  
However, with the available CR data, the propagation uncertainties dominate over the entire energy range as shown in figure~\ref{fig:cs}. 

Upcoming measurements (in particular, from AMS-02~\cite{Incagli:2010}, CALET~\cite{CALET}, and ISS-CREAM~\cite{Cream2014}) are expected to significantly improve our knowledge of propagation parameters and then to reduce the associated uncertainties. 
In that situation, antiproton production cross sections will prevent us to provide predictions for the astrophysical backgrounds as accurate as the forecasted sensitivities.
   
\subsection{The role of charge-dependent solar modulation}\label{sec:charged}

\begin{figure}[!t]
\begin{center}
\includegraphics[width=0.49\textwidth]{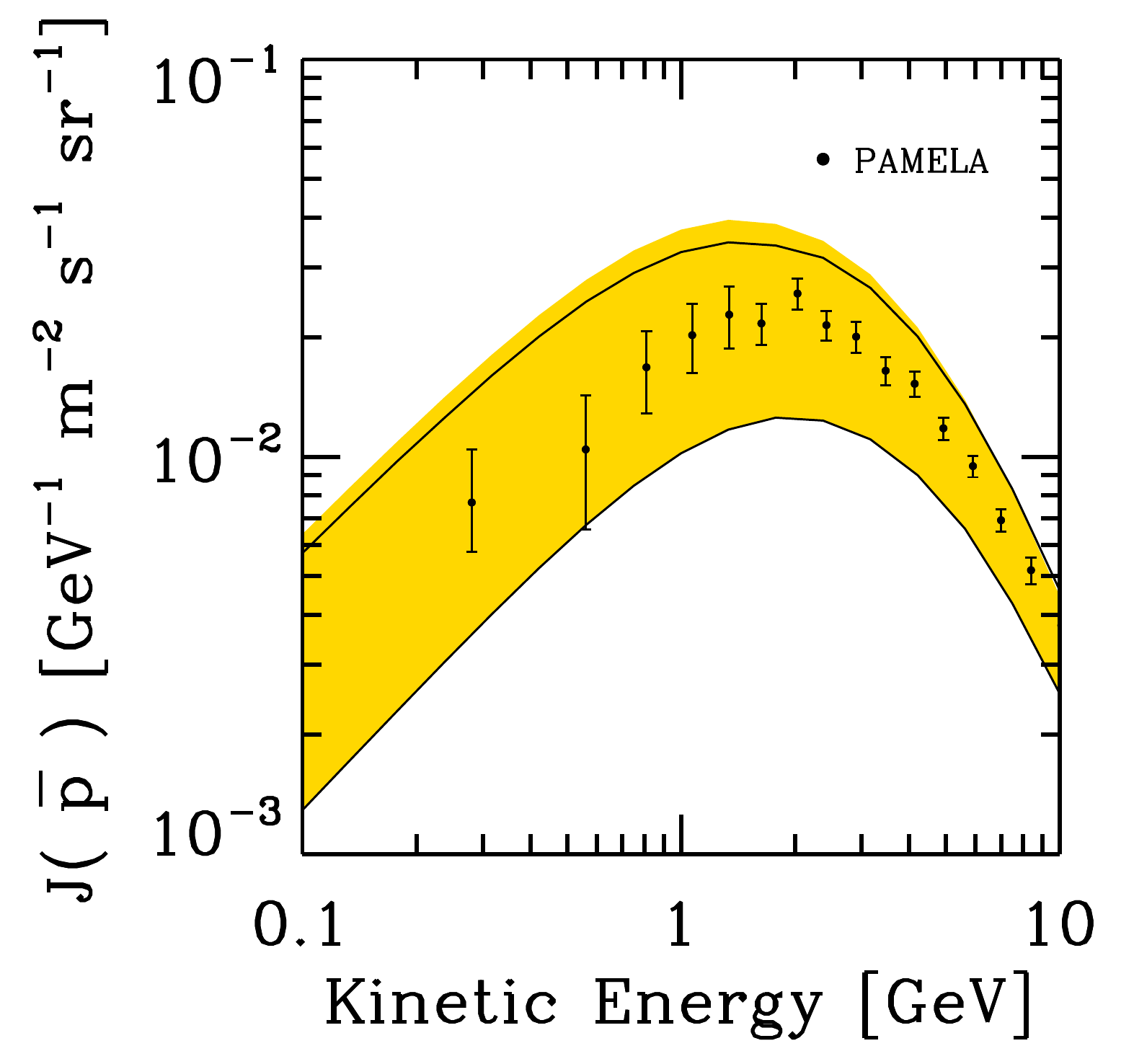}
\caption{The envelope of the secondary antiproton spectra computed with the charge-dependent modulation ({\it black lines}) and compared with that one obtained with the force-field approximation ({\it yellow band}).}
\label{fig:minmax_CD}
\end{center}
\end{figure}

As pointed out in section~\ref{solarmod}, charge-dependent solar modulation can be relevant when the TOA antiproton flux is evaluated. 
Therefore, we compare here our predictions of the extreme fluxes based on the force-field approximation with those obtained with a charge-dependent modulation model.  

To modulate the antiproton flux in the charge-dependent scenario, we develop the following strategy: 

\begin{itemize}

\item
For each propagation model, we consider as free parameters: 1) solar magnetic field polarity; 2) $\alpha$ (HCS tilt angle); 3) $\lambda_0$ (normalization of the parallel mean free path); 4) $\delta$ (power-law slope of the heliosphere diffusion coefficient as function of rigidity). 
Solar polarity and $\alpha$ are fixed by the data-taking period, since they can be obtained by direct measurements~\cite{TiltAngle}, while we determine $\lambda_0$ and $\delta$ by fitting the predicted TOA proton flux against the low-energy PAMELA measurements.

\item
We use the same set of parameters obtained from protons to modulate the LIS antiproton flux.

\end{itemize}

In figure~\ref{fig:minmax_CD} we show the extreme antiproton fluxes as obtained with our charge-dependent modulation model.
We immediately notice that the more detailed treatment of solar modulation does not impact significantly on the determination of the minimum flux, while differences up to $\sim 20 - 30$\% are shown for the maximum one (see also~\cite{Cirelli2014} for a more detailed discussion on this issue).

To understand this result, we point out that the models giving the minimal antiproton TOA fluxes are the ones with the largest LIS proton spectrum and, therefore, they need a stronger modulation to reproduce the data. 
Stronger modulation is associated with a larger Fisk potential or a smaller heliospheric parallel mean free path. 
In this situation, diffusion dominates over charge-dependent drifts and the modulation of particles with different charges becomes equivalent. 

\section{Conservative limit on DM models from antiproton data}

\begin{figure}[!t]
\begin{center}
\includegraphics[width=0.55\textwidth]{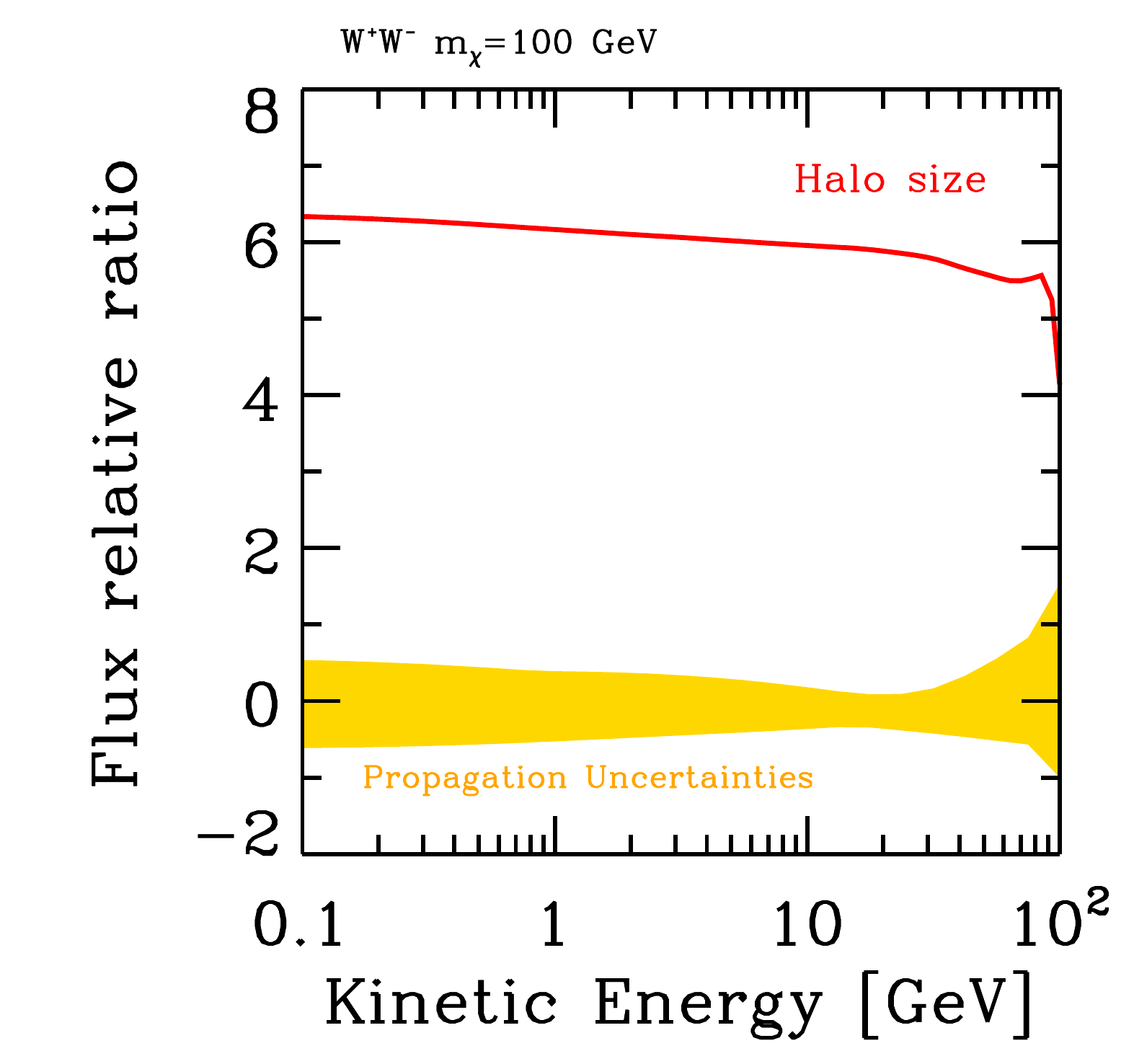}
\caption{Uncertainties on the flux of primary antiprotons originating from DM annihilation due to the propagation parameters (yellow band) and to the halo size (red line). For the latter, the ratio between the two extreme cases $L = 16$ kpc and $L = 2$ kpc is considered}
\label{fig:dmPropOnly}
\end{center}
\end{figure} 

In this section we derive the most conservative -- with respect to all the background uncertainties discussed before -- constraints on the DM annihilation cross section for the DM WIMP scenarios introduced in section~\ref{sec:dm}. 

To propagate DM antiprotons, we choose $L = 2$~kpc since it is the minimum value compatible with synchrotron diffuse emission observations~\cite{DiBernardo:2012zu,Bringmann:2011py}. We note here that, while the actual value of $L$ is irrelevant for the secondary antiprotons (see Fig.~\ref{fig:vsL}), DM antiprotons can change significantly and, in fact, this parameter is the most important one to evaluate this contribution. 
In particular, thinner halos underproduce the DM $\bar{p}$ flux, and therefore $L=2$~kpc corresponds to the minimum flux expected from a given DM model (see~\cite{Evoli:2011id} for a more detailed discussion) and, for that reason, to the less stringent bound. 

In order to quantify these statements, in Fig.~\ref{fig:dmPropOnly} we show the ratio between the antiproton flux corresponding to $L=16$~kpc and $L=2$~kpc, compared to the uncertainty band due to the poor knowledge of the propagation parameters. The plot clearly shows how, as far as primary antiprotons from DM annihilations are concerned, the uncertainty on the halo size is strongly dominant.

We also remark that we can safely neglect the charge-dependent effects in the determination of the minimum background (see section~\ref{sec:charged}).

\begin{figure}[!t]
\begin{center}
\includegraphics[width=0.55\textwidth]{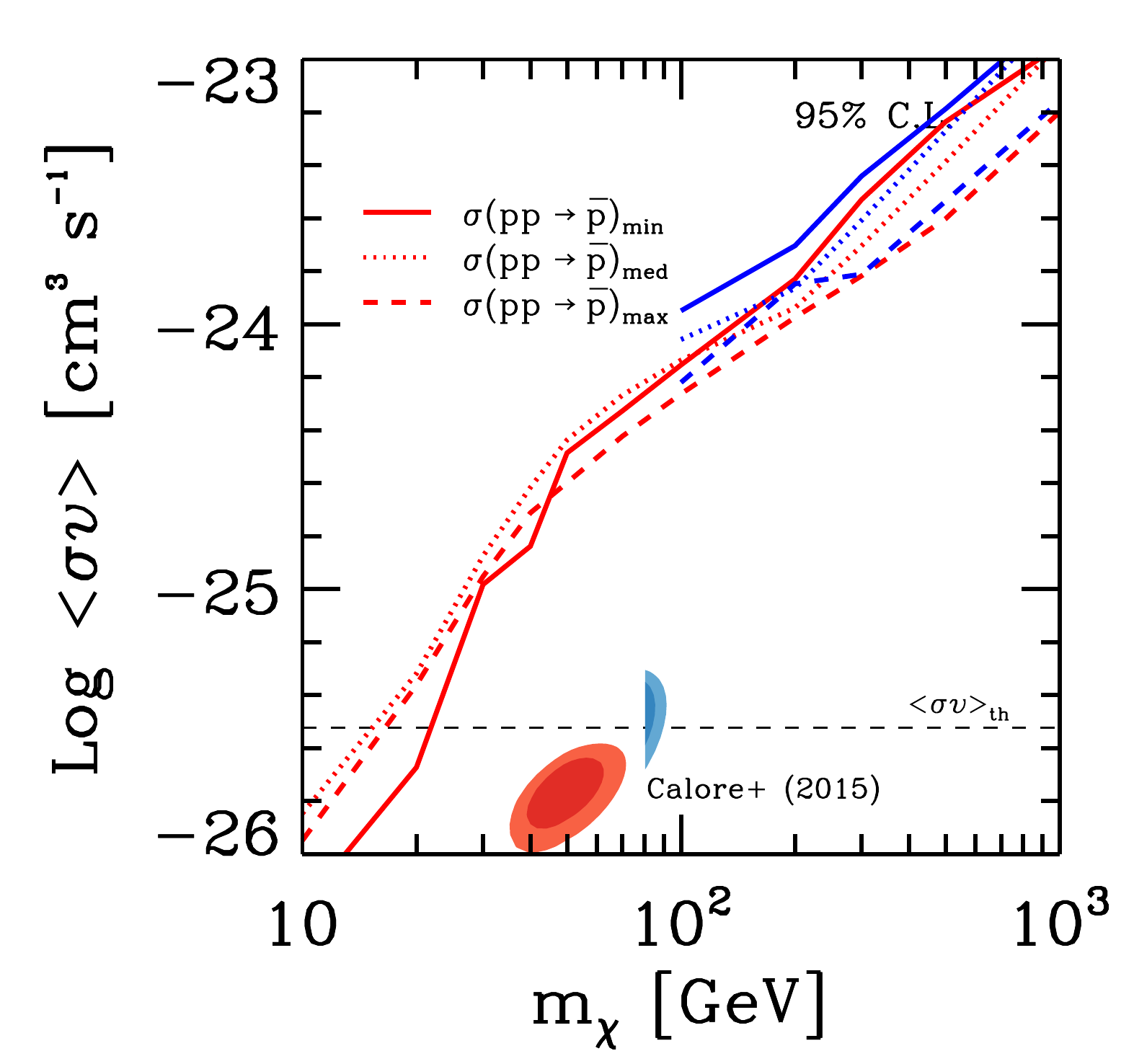}
\caption{Antiproton bounds on DM annihilation rate. {\it Red lines:} $b \bar{b}$ channel for NFW profile for different assumption for the secondary $\bar{p}$ production. {\it Blue lines:} the same for the $WW$ channel. The results obtained with a gNFW profile with $\gamma = 1.2$ are indistinguishable from the NFW ones.} 
\label{fig:bound}
\end{center}
\end{figure} 

In order to get the most conservative bound with respect to the propagation setup, a naive strategy would be the following: Starting from a minimal background,
 the DM component is added until a best fit to the data is reached, and then the DM cross section is increased until the quality of the fit is worsened up to the $2\sigma$ level.
However, although the models providing the minimal background are compatible with the B/C within $3\sigma$, they do not necessarily provide a satisfactory fit of the antiproton flux. 
The addition of a DM component could still leave unexplained the antiproton flux at energies above the DM mass.

Hence, we determine the 2$\sigma$ exclusion contour in the plane ($m_{\rm DM}$, $\langle \sigma v \rangle$) as follows.

For each good propagation model (based on B/C and nuclear data), we first compute the secondary background and the DM flux for a given DM 
mass, 
channel and profile. 
We then find the best-fit value for the annihilation cross section. 

If the combination of background and DM (computed assuming the best-fit cross section) satisfies the following condition:

\begin{equation}
\chi^2 \, (\sigma_{\rm best fit} v) \, \leq \chi^2_{0.05} \, (\mathcal{F} = 23 - 1 = 22) \simeq 33.92 	
\end{equation}

(where the number of degrees of freedom $\mathcal{F}$ is computed taking into account $23$ data points and $1$ parameter), then we retain the model. In other words, we reject
propagation
 models with a p-value lower than $5$\%.

If the model passes this test, we  compute the value of $\langle \sigma v \rangle$ above which the fit worsens beyond the $2\sigma$ level with respect to the best fit. More precisely, we set the following threshold, corresponding again to a p-value equal to $5$\%:

\begin{equation}
\chi^2 \, (\sigma_{\rm max} v) - \chi^2 \, (\sigma_{\rm best fit} v) \, \leq \chi^2_{0.05} \, (\mathcal{F} = 1) \simeq 3.84 \,
\end{equation}

where now the only degree of freedom is the annihilation cross section.

Finally, a scan over the models is performed: The limit value shown in Fig.~$7$ is therefore obtained by taking the maximum cross section (with respect to the different propagation models) satisfying the bound described above for given DM mass.
  
We remark that this strategy allows to obtain the bound by propagating secondary and DM antiprotons consistently with the same propagation model.
    
In the same plot, we compare the limits we obtain when the background is computed with the fiducial model for the $\bar{p}$ production cross-section and with their minimal and maximal values. Because of the strategy adopted to derive them, the case in which the most conservative limits are obtained with the minimal production cross sections is not always prevalent: Indeed, it may happen that larger production cross sections select a different subset of propagation models to reproduce the $\bar{p}$ measurements in combination with DM. Some of the propagation models allowed with fiducial or maximal cross-section, but not with the minimal case, may produce a lower DM $\bar{p}$ flux leading to a larger $2\sigma$ annihilation cross section.

In figure~\ref{fig:bound} the reader can see our results for the maximum allowed annihilation cross section for the $b \bar{b}$ and $W^+W^-$ annihilation channels. 
The maximum allowed cross section we find for $b \bar{b}$ is around one order of magnitude larger than what the authors of~\cite{Cirelli2014} found for the charge-symmetric modulation case. The main difference in our approach is to make use of a broader analysis of the propagation and nuclear uncertainties in order to determine the background.

Our results can be now compared with the DM interpretation of the recently claimed signal in the gamma-ray channel located in the inner few degrees around the GC~\cite{Hooperon2014}.
 
In~\cite{Hooperon2014} the authors show that a DM particle with mass $\sim 43$ GeV annihilating into $b \bar{b}$ with a cross section $\langle\sigma v\rangle \, \simeq \, 2.2 \cdot 10^{-26} \, {\rm cm^3 s^{-1}}$ (for the Inner Galaxy analysis) and distributed according to a gNFW profile with $\gamma = 1.2$ can accomodate the anomalous excess. 

The detailed analysis reported in~\cite{Calore2014} provided a better quantification of the systematic uncertainties affecting the proposed signal; more recently, the Fermi-LAT collaboration released preliminary estimates of the energy spectrum of this excess, based on four qualitatively different background models~\cite{Fermi_GC_analysis}. 
A wide set of DM masses and annihilation channels are compatible with these new analyses (see, e.g.,~\cite{Agrawal2014}). It has been also shown that, in the context of the Minimal Supersymmetric Model framework, these candidates are not in tension with LHC or direct detection constraints.
In figure~\ref{fig:bound} we compare our findings with the favored regions of annihilation cross sections connected to the GC excess as reported in~\cite{Calore2014}.

The bottom line of this analysis can be summarized as it follows: 
Although some fiducial models of CR propagation would produce strong tension with the DM interpretation of the GC excess (see, e.g.,~\cite{Bringmann2014,Cirelli2014}), 
given the large uncertainties on the propagation parameters (for the secondary $\bar{p}$) and on the halo height (for the DM $\bar{p}$), the antiproton channel cannot be invoked to conclusively exclude this hypothesis. 

\section{Discussion on AMS-02 data}

\begin{figure}[!t]
\begin{center}
\includegraphics[width=0.49\textwidth]{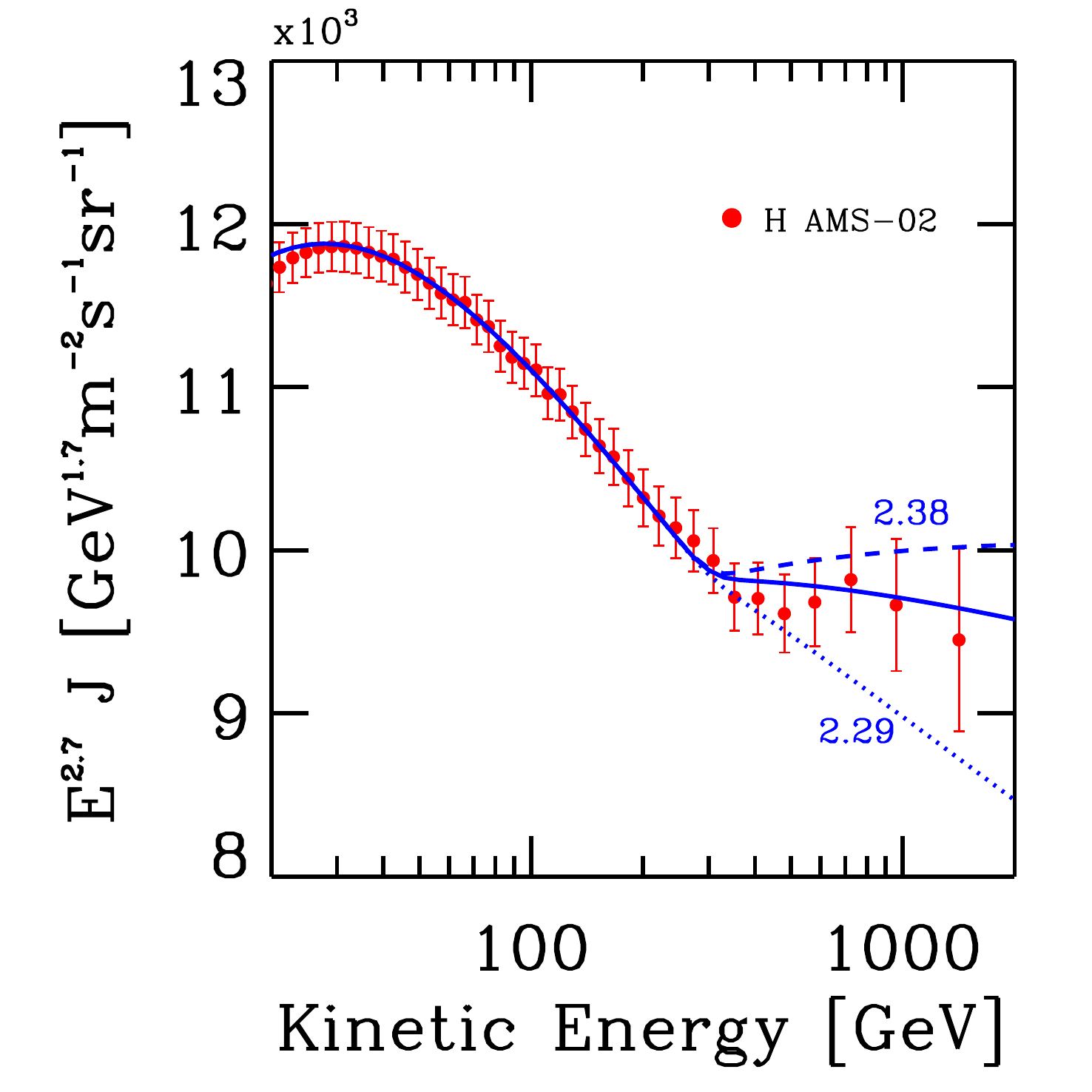}\hspace{\stretch{1}}
\includegraphics[width=0.49\textwidth]{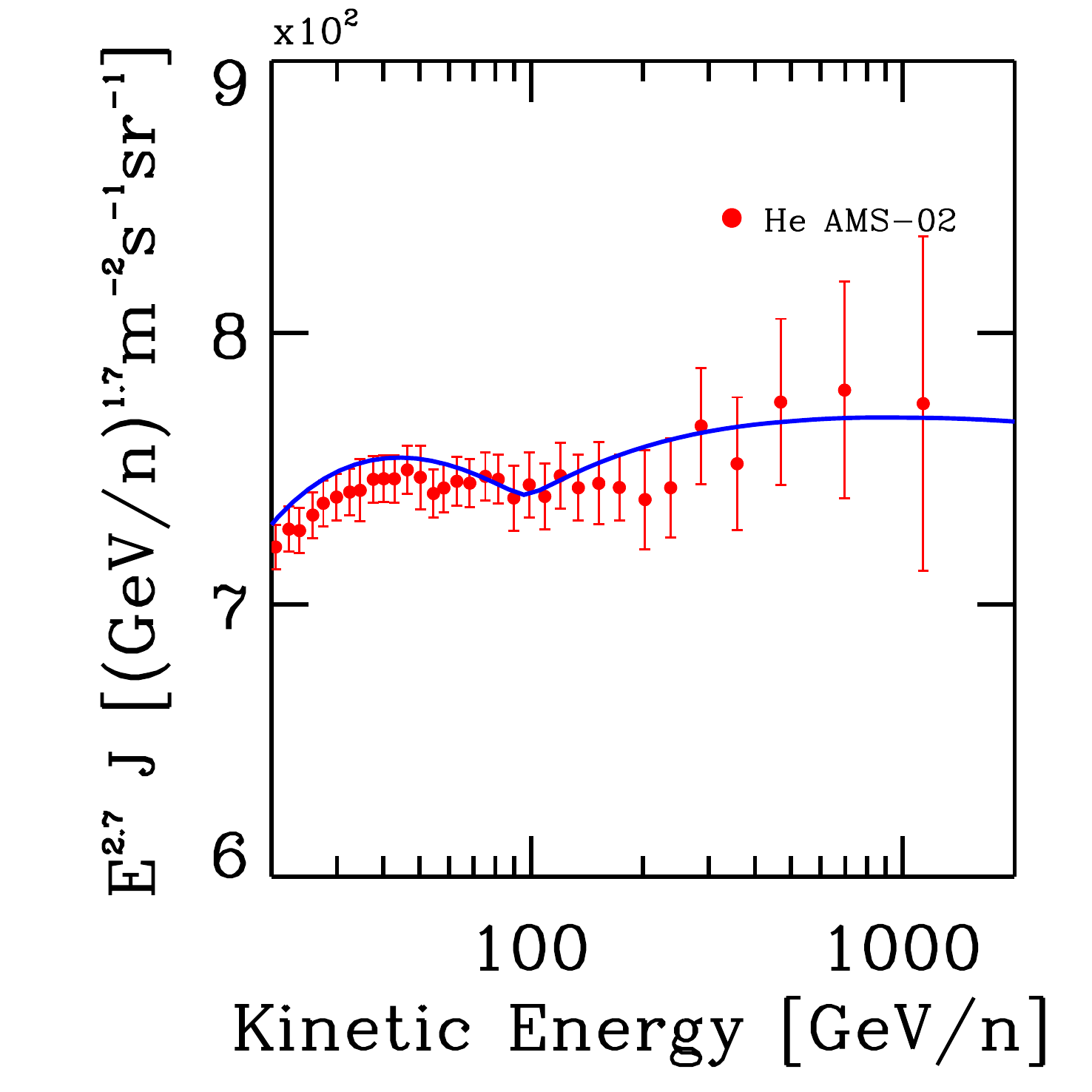}
\caption{Our reference model is compared to AMS-02 proton~\cite{Aguilar:2015} ({\it left plot}) and helium ({\it right plot}) data. With the dotted and dashed lines we show the minimal and maximal breaks compatible with the hardening measured by AMS-02.}
\label{fig:p_ams}
\end{center}
\end{figure}

In this section we focus on the recently released AMS-02 data, including protons~\cite{Aguilar:2015}, and preliminary helium, B/C and $\bar{p}/p$ ratio~\cite{amstalk}, with energy range extending to 450 GeV.

In particular, we take a closer look at the new impressively accurate data on the $\bar{p}/p$ ratio and we attempt to evaluate their compatibility with the other hadronic observables.
Given the preliminary nature of the released data we do not attempt a statistical analysis of the uncertainties associated with propagation. In this perspective, the final release of the secondary/primary measurements, when systematic and statistical errors are fully accounted for, will be crucial.

\begin{figure}[!t]
\begin{center}
\includegraphics[width=0.55\textwidth]{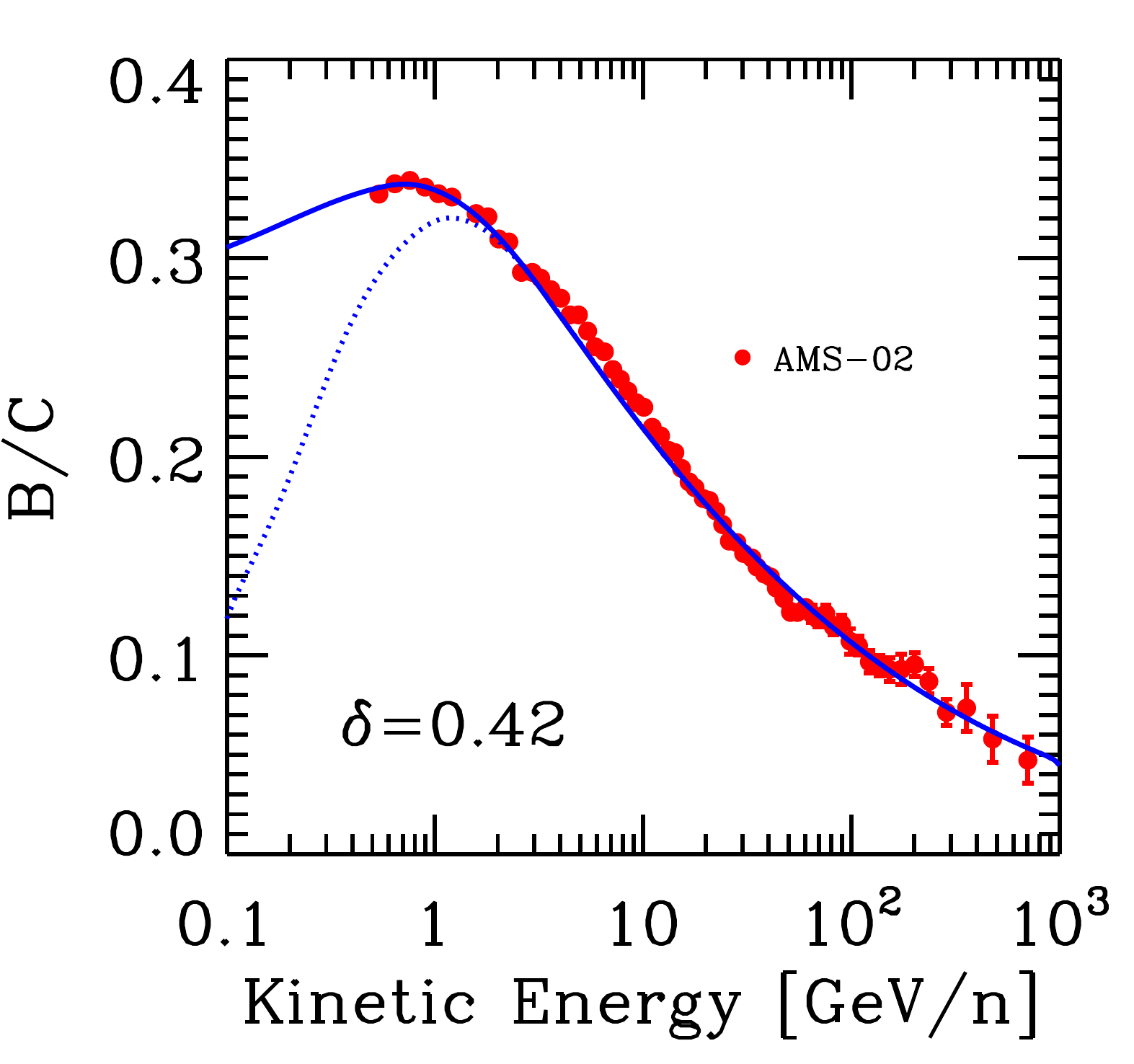}
\caption{Our reference model compared to AMS-02 preliminary B/C data. {\it Solid line:} the TOA spectrum modulated with $\phi = 0.6$~GV; {\it dotted line:} the LIS spectrum.}
\label{fig:bc_ams}
\end{center}
\end{figure}

A propagation model chosen among those considered in section~\ref{sec:selection}, and compatible with preliminary B/C measurements, is shown in figure~\ref{fig:bc_ams}. The propagation parameters are: $D_0 = 1.5$, $\delta = 0.42$, $v_A = 27$, $dV_C/dz = 14$, $\gamma_C = 2.46$, $\gamma_H = 2.44/2.31$.
For comparison, the same value for $\delta$ was found by~\cite{Genolini:2015} using the same datasets.

Remarkably, the predicted B/C ratio reproduces the AMS-02 data over more than three orders of magnitude in energy. It is worth noting here that the $\delta$ required by the new high-energy measurements is in perfect agreement with the best-fit value obtained in our earlier statistical analysis~\cite{DiBernardo10}, based on the available high-energy measurements preceding PAMELA and AMS-02 releases.
 
We also tune the proton and helium injection slopes to accomodate the AMS-02 data. For the protons, we also consider the minimal and maximal injection slopes at high energy compatible with the data.
The reader can see the comparison with the new datasets in figure~\ref{fig:p_ams}. 

Armed with a model fully consistent with all the preliminary nuclear observables, we can finally compare our prediction for the $\bar{p}/p$ ratio with the data.

\begin{figure}[!t]
\begin{center}
\includegraphics[width=0.55\textwidth]{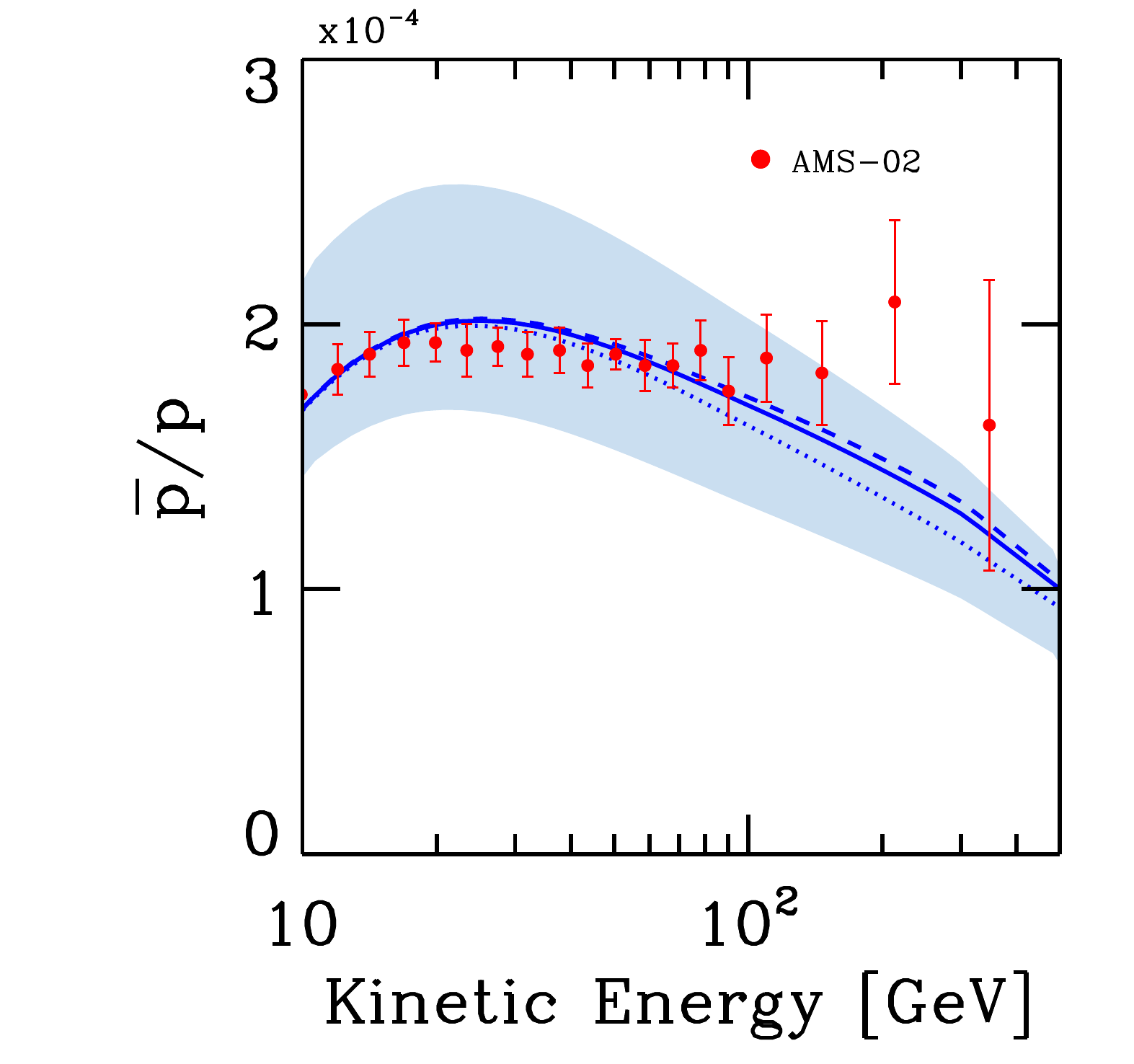}
\caption{Our reference model compared to AMS-02 preliminary $\bar{p}/p$ data. Blue solid line: the $\bar{p}/p$ spectrum computed with the fiducial cross sections from~\cite{Kappl:2014hha}, with the optimal hardening in the proton and helium injection spectra. Dotted and dashed lines: the $\bar{p}/p$ spectrum computed with the minimal and maximal hardening in the proton spectrum as in Fig. \ref{fig:p_ams}. The blue band reports the uncertainty associated to the production cross sections.}
\label{fig:ap_ams}
\end{center}
\end{figure}

In figure~\ref{fig:ap_ams} we show this comparison.
The computation of the secondary flux is performed using the fiducial value of the cross sections provided by~\cite{Kappl:2014hha}, and the associated uncertainty is shown as a blue band. 

We conclude that, even without considering all the relevant uncertainties associated with propagation or injection slopes, our predictions for the $\bar{p}/p$  are in good agreement with the preliminary data in the entire energy range.  
Our findings are then in agreement with the conclusions of \cite{Giesen:2015}, although our analysis relies on the B/C data from the same experiment for the assessment of the propagation model.
 
\section{Conclusions} 

We presented a revisited study of the dominant uncertainties in the determination of the CR secondary antiproton spectrum. 

By performing a scan over the parameter space relevant for CR propagation, we identified a set of models compatible with B/C, proton, helium and carbon data provided by the PAMELA experiment. 
We were then able to bracket the minimum and maximum secondary antiproton fluxes constrained by local observables and we compared the associated uncertainty band with the errors related to the production cross sections. 
It is the first time that such analysis has been performed by using comprehensive numerical simulations of CR propagation in the Galaxy and the Heliosphere. 
More importantly, we used for the first time a complete set of measurements from the same experiment:
Using consistent data from the same data-taking period allowed us to reduce the uncertainties due to solar modulation. 

{Similarly to previous results, we found that the determination of the (almost unknown) diffusion halo height is irrelevant for the computation of the secondary antiproton flux since this parameter is degenerate with the diffusion coefficient normalization $D_0$.}
In addition, we found that using the recent PAMELA data, the uncertainty on the propagation model dominates over the nuclear ones. 

Our result has important implications for the indirect search of primary $\bar{p}$ from DM annihilations in the galactic halo.
Therefore, we provided the most conservative --  with respect to the mentioned effects -- constraints on the annihilation rate for some popular DM models recently investigated in connection to hints of DM signals in other detection channels.

Our method may be taken as a reference procedure to be exploited when the final measurements for all the relevant channels are published by the AMS-02 collaboration.

At the moment, the preliminary release by the AMS-02 collaboration of  nuclear data does not permit to perform a statistical analysis. 
Nevertheless, we found that the model in agreement with AMS-02 proton, helium, and B/C data is compatible with the $\bar{p}/p$ spectrum.
Therefore, we do not report any significant anomaly in this observable.  
Our result is then consistent with the conclusions presented in~\cite{Giesen:2015}.

\section{Acknowledgements}

We thank Torsten Bringmann, Marco Cirelli, Nicola Mori, Pasquale D.~Serpico, Piero Ullio, Alfredo Urbano and Christoph Weniger for useful discussions.
We are indebted to Mattia Di Mauro and Martin Winkler for providing us their cross section models and to Francesca Calore for the confidence regions related to the GC analysis.
Carmelo Evoli acknowledges support from the ``Helmholtz Alliance for Astroparticle Physics HAP'' funded by the Initiative and Networking Fund of the Helmholtz Association.
Daniele Gaggero acknowledges the SFB 676 research fellowship from the University of Hamburg as well as the hospitality of DESY.

\bibliographystyle{JHEP}
\bibliography{exoticap}

\providecommand{\href}[2]{#2}\begingroup\raggedright\begin{thebibliography}{10}

\bibitem{Incagli:2010}
M.~{Incagli}, {\it {Astroparticle Physiscs with AMS02}},  in {\em American
  Institute of Physics Conference Series} (C.~{Cecchi}, S.~{Ciprini},
  P.~{Lubrano}, and G.~{Tosti}, eds.), vol.~1223 of {\em American Institute of
  Physics Conference Series}, pp.~43--49, Mar., 2010.

\bibitem{Lavalle:2012ef}
J.~Lavalle and P.~Salati, {\it {Dark Matter Indirect Signatures}},  {\em
  Comptes Rendus Physique} {\bf 13} (2012) 740--782,
  [\href{http://arxiv.org/abs/1205.1004}{{\tt arXiv:1205.1004}}].

\bibitem{Silk:1984}
J.~{Silk} and M.~{Srednicki}, {\it {Cosmic-ray antiprotons as a probe of a
  photino-dominated universe}},  {\em Physical Review Letters} {\bf 53} (Aug.,
  1984) 624--627.

\bibitem{Maurin:2001}
D.~{Maurin}, F.~{Donato}, R.~{Taillet}, and P.~{Salati}, {\it {Cosmic Rays
  below Z=30 in a Diffusion Model: New Constraints on Propagation Parameters}},
   {\em \apj} {\bf 555} (July, 2001) 585--596,
  [\href{http://arxiv.org/abs/astro-ph/0101231}{{\tt astro-ph/0101231}}].

\bibitem{Moskalenko02}
I.~V. {Moskalenko}, A.~W. {Strong}, J.~F. {Ormes}, and M.~S. {Potgieter}, {\it
  {Secondary Antiprotons and Propagation of Cosmic Rays in the Galaxy and
  Heliosphere}},  {\em \apj} {\bf 565} (Jan., 2002) 280--296,
  [\href{http://arxiv.org/abs/astro-ph/0106567}{{\tt astro-ph/0106567}}].

\bibitem{DiBernardo10}
G.~{di Bernardo}, C.~{Evoli}, D.~{Gaggero}, D.~{Grasso}, and L.~{Maccione},
  {\it {Unified interpretation of cosmic ray nuclei and antiproton recent
  measurements}},  {\em Astroparticle Physics} {\bf 34} (Dec., 2010) 274--283,
  [\href{http://arxiv.org/abs/0909.4548}{{\tt arXiv:0909.4548}}].

\bibitem{Putze:2010}
A.~{Putze}, L.~{Derome}, and D.~{Maurin}, {\it {A Markov Chain Monte Carlo
  technique to sample transport and source parameters of Galactic cosmic rays.
  II. Results for the diffusion model combining B/C and radioactive nuclei}},
  {\em \aap} {\bf 516} (June, 2010) A66,
  [\href{http://arxiv.org/abs/1001.0551}{{\tt arXiv:1001.0551}}].

\bibitem{Trotta:2011}
R.~{Trotta}, G.~{J{\'o}hannesson}, I.~V. {Moskalenko}, T.~A. {Porter}, R.~{Ruiz
  de Austri}, and A.~W. {Strong}, {\it {Constraints on Cosmic-ray Propagation
  Models from A Global Bayesian Analysis}},  {\em \apj} {\bf 729} (Mar., 2011)
  106, [\href{http://arxiv.org/abs/1011.0037}{{\tt arXiv:1011.0037}}].

\bibitem{Cholis2013}
I.~{Cholis} and D.~{Hooper}, {\it {Constraining the origin of the rising cosmic
  ray positron fraction with the boron-to-carbon ratio}},  {\em \prd} {\bf 89}
  (Feb., 2014) 043013, [\href{http://arxiv.org/abs/1312.2952}{{\tt
  arXiv:1312.2952}}].

\bibitem{Mertsch2014}
P.~{Mertsch} and S.~{Sarkar}, {\it {AMS-02 data confront acceleration of cosmic
  ray secondaries in nearby sources}},  {\em \prd} {\bf 90} (Sept., 2014)
  061301, [\href{http://arxiv.org/abs/1402.0855}{{\tt arXiv:1402.0855}}].

\bibitem{Adriani:2011cu}
{\bf PAMELA Collaboration} Collaboration, O.~Adriani et~al., {\it {PAMELA
  Measurements of Cosmic-ray Proton and Helium Spectra}},  {\em Science} {\bf
  332} (2011) 69--72, [\href{http://arxiv.org/abs/1103.4055}{{\tt
  arXiv:1103.4055}}].

\bibitem{Adriani:2014xoa}
O.~{Adriani} et~al., {\it {Measurement of Boron and Carbon Fluxes in Cosmic
  Rays with the PAMELA Experiment}},  {\em The Astrophysical Journal} {\bf 791}
  (Aug., 2014) 93, [\href{http://arxiv.org/abs/1407.1657}{{\tt
  arXiv:1407.1657}}].

\bibitem{Evoli:2011id}
C.~Evoli, I.~Cholis, D.~Grasso, L.~Maccione, and P.~Ullio, {\it {Antiprotons
  from dark matter annihilation in the Galaxy: astrophysical uncertainties}},
  {\em Phys.Rev.} {\bf D85} (2012) 123511,
  [\href{http://arxiv.org/abs/1108.0664}{{\tt arXiv:1108.0664}}].

\bibitem{Kappl:2014hha}
R.~{Kappl} and M.~W. {Winkler}, {\it {The cosmic ray antiproton background for
  AMS-02}},  {\em \jcap} {\bf 9} (Sept., 2014) 51,
  [\href{http://arxiv.org/abs/1408.0299}{{\tt arXiv:1408.0299}}].

\bibitem{NA49}
``{NA49 experiment website}.'' \url{http://na49info.web.cern.ch/na49info/}.

\bibitem{Maccione:2012cu}
L.~Maccione, {\it {Low energy cosmic ray positron fraction explained by
  charge-sign dependent solar modulation}},  {\em Phys.Rev.Lett.} {\bf 110}
  (2013) 081101, [\href{http://arxiv.org/abs/1211.6905}{{\tt
  arXiv:1211.6905}}].

\bibitem{Donato:2003}
F.~Donato, N.~Fornengo, D.~Maurin, and P.~Salati, {\it {Antiprotons in cosmic
  rays from neutralino annihilation}},  {\em Phys. Rev.} {\bf D69} (2004)
  063501, [\href{http://arxiv.org/abs/astro-ph/0306207}{{\tt
  astro-ph/0306207}}].

\bibitem{Donato:2008jk}
F.~Donato, D.~Maurin, P.~Brun, T.~Delahaye, and P.~Salati, {\it {Constraints on
  WIMP Dark Matter from the High Energy PAMELA $\bar{p}/p$ data}},  {\em
  Phys.Rev.Lett.} {\bf 102} (2009) 071301,
  [\href{http://arxiv.org/abs/0810.5292}{{\tt arXiv:0810.5292}}].

\bibitem{Bringmann:2006im}
T.~Bringmann and P.~Salati, {\it {The galactic antiproton spectrum at high
  energies: background expectation vs. exotic contributions}},  {\em Phys.
  Rev.} {\bf D75} (2007) 083006,
  [\href{http://arxiv.org/abs/astro-ph/0612514}{{\tt astro-ph/0612514}}].

\bibitem{AMS_preliminary}
``{Preliminary AMS data}.'' \url{http://www.ams02.org/2015/04/}.

\bibitem{Evoli:2008dv}
C.~Evoli, D.~Gaggero, D.~Grasso, and L.~Maccione, {\it {Cosmic-Ray Nuclei,
  Antiprotons and Gamma-rays in the Galaxy: a New Diffusion Model}},  {\em
  JCAP} {\bf 0810} (2008) 018, [\href{http://arxiv.org/abs/0807.4730}{{\tt
  arXiv:0807.4730}}].

\bibitem{Dragonweb}
``{DRAGON code}.'' \url{http://www.dragonproject.org/}.

\bibitem{Moskalenko1998}
I.~V. {Moskalenko}, A.~W. {Strong}, and O.~{Reimer}, {\it {Diffuse galactic
  gamma rays, cosmic-ray nucleons and antiprotons}},  {\em \aap} {\bf 338}
  (Oct., 1998) L75--L78, [\href{http://arxiv.org/abs/astro-ph/9808084}{{\tt
  astro-ph/9808084}}].

\bibitem{DiBernardo:2012zu}
G.~{Di Bernardo}, C.~{Evoli}, D.~{Gaggero}, D.~{Grasso}, and L.~{Maccione},
  {\it {Cosmic ray electrons, positrons and the synchrotron emission of the
  Galaxy: consistent analysis and implications}},  {\em JCAP} {\bf 3} (Mar.,
  2013) 36, [\href{http://arxiv.org/abs/1210.4546}{{\tt arXiv:1210.4546}}].

\bibitem{Lavalle:2014kca}
J.~{Lavalle}, D.~{Maurin}, and A.~{Putze}, {\it {Direct constraints on
  diffusion models from cosmic-ray positron data: Excluding the minimal model
  for dark matter searches}},  {\em \prd} {\bf 90} (Oct., 2014) 081301,
  [\href{http://arxiv.org/abs/1407.2540}{{\tt arXiv:1407.2540}}].

\bibitem{Gaggero2013}
D.~{Gaggero}, L.~{Maccione}, G.~{Di Bernardo}, C.~{Evoli}, and D.~{Grasso},
  {\it {Three-Dimensional Model of Cosmic-Ray Lepton Propagation Reproduces
  Data from the Alpha Magnetic Spectrometer on the International Space
  Station}},  {\em Physical Review Letters} {\bf 111} (July, 2013) 021102,
  [\href{http://arxiv.org/abs/1304.6718}{{\tt arXiv:1304.6718}}].

\bibitem{Kissmann:2015}
R.~{Kissmann}, M.~{Werner}, O.~{Reimer}, and A.~W. {Strong}, {\it {Propagation
  in 3D spiral-arm cosmic-ray source distribution models and secondary particle
  production using PICARD}},  {\em Astroparticle Physics} {\bf 70} (Oct., 2015)
  39--53, [\href{http://arxiv.org/abs/1504.08249}{{\tt arXiv:1504.08249}}].

\bibitem{Ptuskin06}
V.~S. {Ptuskin}, I.~V. {Moskalenko}, F.~C. {Jones}, A.~W. {Strong}, and V.~N.
  {Zirakashvili}, {\it {Dissipation of Magnetohydrodynamic Waves on Energetic
  Particles: Impact on Interstellar Turbulence and Cosmic-Ray Transport}},
  {\em \apj} {\bf 642} (May, 2006) 902--916,
  [\href{http://arxiv.org/abs/astro-ph/0510335}{{\tt astro-ph/0510335}}].

\bibitem{EvoliYan2014}
C.~{Evoli} and H.~{Yan}, {\it {Cosmic Ray Propagation in Galactic Turbulence}},
   {\em \apj} {\bf 782} (Feb., 2014) 36,
  [\href{http://arxiv.org/abs/1310.5732}{{\tt arXiv:1310.5732}}].

\bibitem{Cesarsky}
C.~J. {Cesarsky}, {\it {Cosmic-ray confinement in the galaxy}},  {\em \araa}
  {\bf 18} (1980) 289--319.

\bibitem{Blasi:2012yr}
P.~Blasi, E.~Amato, and P.~D. Serpico, {\it {Spectral breaks as a signature of
  cosmic ray induced turbulence in the Galaxy}},  {\em Phys.Rev.Lett.} {\bf
  109} (2012) 061101, [\href{http://arxiv.org/abs/1207.3706}{{\tt
  arXiv:1207.3706}}].

\bibitem{Donato01}
F.~{Donato}, D.~{Maurin}, P.~{Salati}, A.~{Barrau}, G.~{Boudoul}, and
  R.~{Taillet}, {\it {Antiprotons from Spallations of Cosmic Rays on
  Interstellar Matter}},  {\em \apj} {\bf 563} (Dec., 2001) 172--184,
  [\href{http://arxiv.org/abs/astro-ph/0103150}{{\tt astro-ph/0103150}}].

\bibitem{Everett}
J.~E. {Everett}, E.~G. {Zweibel}, R.~A. {Benjamin}, D.~{McCammon}, L.~{Rocks},
  and J.~S. {Gallagher}, III, {\it {The Milky Way's Kiloparsec-Scale Wind: A
  Hybrid Cosmic-Ray and Thermally Driven Outflow}},  {\em \apj} {\bf 674}
  (Feb., 2008) 258--270, [\href{http://arxiv.org/abs/0710.3712}{{\tt
  arXiv:0710.3712}}].

\bibitem{Breitschwerdt}
D.~{Breitschwerdt}, {\it {Astrophysics: Blown away by cosmic rays}},  {\em
  \nat} {\bf 452} (Apr., 2008) 826--827.

\bibitem{Strong:2007nh}
A.~W. Strong, I.~V. Moskalenko, and V.~S. Ptuskin, {\it {Cosmic-ray propagation
  and interactions in the Galaxy}},  {\em Ann.Rev.Nucl.Part.Sci.} {\bf 57}
  (2007) 285--327, [\href{http://arxiv.org/abs/astro-ph/0701517}{{\tt
  astro-ph/0701517}}].

\bibitem{Ferriere:2001rg}
K.~M. Ferriere, {\it {The Interstellar Environment of our Galaxy}},  {\em Rev.
  Mod. Phys.} {\bf 73} (2001) 1031--1066,
  [\href{http://arxiv.org/abs/astro-ph/0106359}{{\tt astro-ph/0106359}}].

\bibitem{1982PhRvD..26.1179T}
L.~C. {Tan} and L.~K. {Ng}, {\it {Parameterization of invariant cross section
  in p-p collisions using a new scaling variable}},  {\em \prd} {\bf 26}
  (Sept., 1982) 1179--1182.

\bibitem{diMauro2014}
M.~{di Mauro}, F.~{Donato}, A.~{Goudelis}, and P.~D. {Serpico}, {\it {New
  evaluation of the antiproton production cross section for cosmic ray
  studies}},  {\em \prd} {\bf 90} (Oct., 2014) 085017,
  [\href{http://arxiv.org/abs/1408.0288}{{\tt arXiv:1408.0288}}].

\bibitem{Cirelli2014}
M.~{Cirelli}, D.~{Gaggero}, G.~{Giesen}, M.~{Taoso}, and A.~{Urbano}, {\it
  {Antiproton constraints on the GeV gamma-ray excess: a comprehensive
  analysis}},  {\em \jcap} {\bf 12} (Dec., 2014) 45,
  [\href{http://arxiv.org/abs/1407.2173}{{\tt arXiv:1407.2173}}].

\bibitem{NFW1996}
J.~F. {Navarro}, C.~S. {Frenk}, and S.~D.~M. {White}, {\it {The Structure of
  Cold Dark Matter Halos}},  {\em \apj} {\bf 462} (May, 1996) 563,
  [\href{http://arxiv.org/abs/astro-ph/9508025}{{\tt astro-ph/9508025}}].

\bibitem{Catena2012}
R.~{Catena} and P.~{Ullio}, {\it {The local dark matter phase-space density and
  impact on WIMP direct detection}},  {\em \jcap} {\bf 5} (May, 2012) 5,
  [\href{http://arxiv.org/abs/1111.3556}{{\tt arXiv:1111.3556}}].

\bibitem{Hooperon2014}
T.~{Daylan}, D.~P. {Finkbeiner}, D.~{Hooper}, T.~{Linden}, S.~K.~N. {Portillo},
  N.~L. {Rodd}, and T.~R. {Slatyer}, {\it {The Characterization of the
  Gamma-Ray Signal from the Central Milky Way: A Compelling Case for
  Annihilating Dark Matter}},  {\em arXiv:1402.6703} (Feb., 2014)
  [\href{http://arxiv.org/abs/1402.6703}{{\tt arXiv:1402.6703}}].

\bibitem{Cirelli2010}
M.~{Cirelli}, G.~{Corcella}, A.~{Hektor}, G.~{H{\"u}tsi}, M.~{Kadastik},
  P.~{Panci}, M.~{Raidal}, F.~{Sala}, and A.~{Strumia}, {\it {PPPC 4 DM ID: a
  poor particle physicist cookbook for dark matter indirect detection}},  {\em
  \jcap} {\bf 3} (Mar., 2011) 51, [\href{http://arxiv.org/abs/1012.4515}{{\tt
  arXiv:1012.4515}}].

\bibitem{Ciafaloni2010}
P.~{Ciafaloni}, D.~{Comelli}, A.~{Riotto}, F.~{Sala}, A.~{Strumia}, and
  A.~{Urbano}, {\it {Weak corrections are relevant for dark matter indirect
  detection}},  {\em \jcap} {\bf 3} (Mar., 2011) 19,
  [\href{http://arxiv.org/abs/1009.0224}{{\tt arXiv:1009.0224}}].

\bibitem{Gleeson68}
L.~J. {Gleeson} and W.~I. {Axford}, {\it {Solar Modulation of Galactic Cosmic
  Rays}},  {\em \apj} {\bf 154} (Dec., 1968) 1011--+.

\bibitem{Strauss:2011}
R.~D. {Strauss}, M.~S. {Potgieter}, A.~{Kopp}, and I.~{B{\"u}sching}, {\it {On
  the propagation times and energy losses of cosmic rays in the heliosphere}},
  {\em Journal of Geophysical Research (Space Physics)} {\bf 116} (Dec., 2011)
  12105.

\bibitem{Strauss:2012}
R.~D. {Strauss}, M.~S. {Potgieter}, I.~{B{\"u}sching}, and A.~{Kopp}, {\it
  {Modelling heliospheric current sheet drift in stochastic cosmic ray
  transport models}},  {\em Astrophysics and Space Science} {\bf 339} (June,
  2012) 223--236.

\bibitem{Giacalone1999}
J.~{Giacalone} and J.~R. {Jokipii}, {\it {The Transport of Cosmic Rays across a
  Turbulent Magnetic Field}},  {\em \apj} {\bf 520} (July, 1999) 204--214.

\bibitem{Cream2014}
E.~S. {Seo}, T.~{Anderson}, D.~{Angelaszek}, S.~J. {Baek}, J.~{Baylon},
  M.~{Bu{\'e}nerd}, M.~{Copley}, S.~{Coutu}, L.~{Derome}, B.~{Fields},
  M.~{Gupta}, J.~H. {Han}, I.~J. {Howley}, H.~G. {Huh}, Y.~S. {Hwang}, H.~J.
  {Hyun}, I.~S. {Jeong}, D.~H. {Kah}, K.~H. {Kang}, D.~Y. {Kim}, H.~J. {Kim},
  K.~C. {Kim}, M.~H. {Kim}, K.~{Kwashnak}, J.~{Lee}, M.~H. {Lee}, J.~T. {Link},
  L.~{Lutz}, A.~{Malinin}, A.~{Menchaca-Rocha}, J.~W. {Mitchell}, S.~{Nutter},
  O.~{Ofoha}, H.~{Park}, I.~H. {Park}, J.~M. {Park}, P.~{Patterson}, J.~R.
  {Smith}, J.~{Wu}, and Y.~S. {Yoon}, {\it {Cosmic Ray Energetics And Mass for
  the International Space Station (ISS-CREAM)}},  {\em Advances in Space
  Research} {\bf 53} (May, 2014) 1451--1455.

\bibitem{Tomassetti:2012}
N.~{Tomassetti} and F.~{Donato}, {\it {Secondary cosmic-ray nuclei from
  supernova remnants and constraints on the propagation parameters}},  {\em
  \aap} {\bf 544} (Aug., 2012) A16, [\href{http://arxiv.org/abs/1203.6094}{{\tt
  arXiv:1203.6094}}].

\bibitem{Genolini:2015}
Y.~{Genolini}, A.~{Putze}, P.~{Salati}, and P.~D. {Serpico}, {\it {Theoretical
  uncertainties in extracting cosmic-ray diffusion parameters: the
  boron-to-carbon ratio}},  {\em \aap} {\bf 580} (Aug., 2015) A9,
  [\href{http://arxiv.org/abs/1504.03134}{{\tt arXiv:1504.03134}}].

\bibitem{Gaggero:2014}
D.~{Gaggero}, A.~{Urbano}, M.~{Valli}, and P.~{Ullio}, {\it {Gamma-ray sky
  points to radial gradients in cosmic-ray transport}},  {\em \prd} {\bf 91}
  (Apr., 2015) 083012, [\href{http://arxiv.org/abs/1411.7623}{{\tt
  arXiv:1411.7623}}].

\bibitem{Gaggero:2015}
D.~{Gaggero}, D.~{Grasso}, A.~{Marinelli}, A.~{Urbano}, and M.~{Valli}, {\it
  {The gamma-ray and neutrino sky: a consistent picture of Fermi-LAT, H.E.S.S.,
  Milagro, and IceCube results}},  {\em arXiv:1504.00227} (Apr., 2015)
  [\href{http://arxiv.org/abs/1504.00227}{{\tt arXiv:1504.00227}}].

\bibitem{CALET}
{\bf CALET} Collaboration, P.~Marrocchesi, {\it {CALET: A calorimeter-based
  orbital observatory for high energy astroparticle physics}},  {\em
  Nucl.Instrum.Meth.} {\bf A692} (2012) 240--245.

\bibitem{TiltAngle}
``{The Wilcox Solar Observatory}.'' \url{http://wso.stanford.edu/}.

\bibitem{Bringmann:2011py}
T.~{Bringmann}, F.~{Donato}, and R.~A. {Lineros}, {\it {Radio data and
  synchrotron emission in consistent cosmic ray models}},  {\em \jcap} {\bf 1}
  (Jan., 2012) 49, [\href{http://arxiv.org/abs/1106.4821}{{\tt
  arXiv:1106.4821}}].

\bibitem{Calore2014}
F.~{Calore}, I.~{Cholis}, and C.~{Weniger}, {\it {Background model systematics
  for the Fermi GeV excess}},  {\em \jcap} {\bf 3} (Mar., 2015) 38,
  [\href{http://arxiv.org/abs/1409.0042}{{\tt arXiv:1409.0042}}].

\bibitem{Fermi_GC_analysis}
``{Fermi GCE preliminary analysis}.''
  \url{http://fermi.gsfc.nasa.gov/science/mtgs/symposia/2014/program/08_Murgia.pdf}.

\bibitem{Agrawal2014}
P.~{Agrawal}, B.~{Batell}, P.~J. {Fox}, and R.~{Harnik}, {\it {WIMPs at the
  galactic center}},  {\em \jcap} {\bf 5} (May, 2015) 11,
  [\href{http://arxiv.org/abs/1411.2592}{{\tt arXiv:1411.2592}}].

\bibitem{Bringmann2014}
T.~{Bringmann}, M.~{Vollmann}, and C.~{Weniger}, {\it {Updated cosmic-ray and
  radio constraints on light dark matter: Implications for the GeV gamma-ray
  excess at the Galactic Center}},  {\em \prd} {\bf 90} (Dec., 2014) 123001,
  [\href{http://arxiv.org/abs/1406.6027}{{\tt arXiv:1406.6027}}].

\bibitem{Aguilar:2015}
M.~{Aguilar}, D.~{Aisa}, B.~{Alpat}, A.~{Alvino}, G.~{Ambrosi}, K.~{Andeen},
  L.~{Arruda}, N.~{Attig}, P.~{Azzarello}, A.~{Bachlechner}, F.~{Barao},
  A.~{Barrau}, L.~{Barrin}, A.~{Bartoloni}, L.~{Basara}, M.~{Battarbee},
  R.~{Battiston}, J.~{Bazo}, U.~{Becker}, M.~{Behlmann}, and et~al., {\it
  {Precision Measurement of the Proton Flux in Primary Cosmic Rays from
  Rigidity 1 GV to 1.8 TV with the Alpha Magnetic Spectrometer on the
  International Space Station}},  {\em Physical Review Letters} {\bf 114} (May,
  2015) 171103.

\bibitem{amstalk}
``{Ams-02 Collaboration, Talks at the ‘AMS Days at CERN’, 15-17 april
  2015}.'' \url{https://indico.cern.ch/event/381134/}.

\bibitem{Giesen:2015}
G.~{Giesen}, M.~{Boudaud}, Y.~{G{\'e}nolini}, V.~{Poulin}, M.~{Cirelli},
  P.~{Salati}, and P.~D. {Serpico}, {\it {AMS-02 antiprotons, at last!
  Secondary astrophysical component and immediate implications for Dark
  Matter}},  {\em \jcap} {\bf 9} (Sept., 2015) 23,
  [\href{http://arxiv.org/abs/1504.04276}{{\tt arXiv:1504.04276}}].

\end{thebibliography}\endgroup

\end{document}